\documentclass[ reprint,
 amsmath,amssymb,
 aps,
]{revtex4-1}

\usepackage{graphicx}
\usepackage{dcolumn}
\usepackage{bm}

\usepackage{ulem}
\usepackage{color}

\begin{document}


\title{Rigorously solvable model for the electrical conductivity of dispersions of hard-core--penetrable-shell particles
and its applications}

\author{M.~Ya.~Sushko}
\email{mrs@onu.edu.ua}
\author{A.~K.~Semenov}
\affiliation{Department of Theoretical Physics and Astronomy,
Mechnikov National University, 2 Dvoryanska St., Odesa 65026,
Ukraine}

\date{\today}

\begin{abstract}
We generalize the compact group approach to conducting systems to
give a self-consistent analytical solution to the problem of the
effective quasistatic electrical conductivity of macroscopically
homogeneous and isotropic dispersions of
hard-core--penetrable-shell particles. The shells are in general
inhomogeneous and characterized by a radially-symmetrical,
piecewise-continuous conductivity profile. The local value of the
conductivity is determined by the shortest distance from the point
of interest to the nearest particle. The effective conductivity is
expressed in terms of the constituents' conductivities and volume
concentrations; the latter account for the statistical
microstructure of the system. The theory effectively incorporates
many-particle effects and is expected to be rigorous in the static
limit. Using the well-tested statistical physics results for the
shell volume concentration, this conclusion is backed up by
mapping the theory on available 3D random resistor network
simulations for hard spheres coated with fully penetrable
concentric shells. Finally, the theory is shown to fit
experimental data for real composite solid electrolytes. The
fitting results indicate that the effect of enhanced electrical
conduction is generally contributed to by several mechanisms.
These are effectively taken into account through the shell
conductivity profile.

\end{abstract}

\maketitle

\section{\label{sec:intro} Introduction}

The objectives of this paper are threefold: (1) to develop  a
homogenization theory for the effective quasistatic electrical
conductivity $\sigma_{\rm eff}$ of macroscopically homogeneous and
isotropic particulate substances and dispersions of particles with
the core-shell morphology; (2)~to test the  theory by comparing
its predictions with available results of random resistor network
(RRN) simulations; and (3) to exemplify the applicability of the
theory to real systems by processing experimental data for
composite solid electrolytes (CSEs) prepared by dispersing fine
insulating particles into matrix ionic conductors.

The indicated class of composites attracts a special attention due
to nontrivial behavior of their $\sigma_{\rm eff}$. Through the
addition of filler particles (for instance, alumina particles,
with electrical conductivity $\sigma_1 \sim 10^{-14} \, {\rm
{S/cm}}$) to matrix ionic conductors (such as polycrystalline
metal halides, whose typical electrical conductivities $\sigma_0
\sim 10^{-10} \div 10^{-5}\, {\rm {S/cm}}$), $\sigma_{\rm {eff}}$
of the resulting CSEs can be increased dramatically,
by one to three orders of magnitude
as compared to $\sigma_0$. This effect is called enhanced ionic
conduction. Since its discovery by Liang~\cite{Liang1973} in
polycrystalline lithium iodide containing alumina particles, it
has been observed in dozens of CSEs (for a detailed bibliography,
see reviews \cite{Wagner1980,
Takahashi1989,Dudney1989,Maier1995,Agrawal1999,Uvarov2011,
Kharton2011,Gao2016}) and composite polymer-based electrolytes
(\cite{Knauth2008,Sequeira2010}) to make  those promising
materials for electrolytic applications.

Experiment also reveals that the maximum conduction enhancement
usually occurs as the  filler volume concentration $c$ reaches
values in between 0.1 and 0.4. It is followed by a decrease in
$\sigma_{\rm {eff}}$ as $c$ is further increased. Such a
nonmonotonic dependence of $\sigma_{\rm {eff}}$ upon $c$ is a
challenging problem for homogenization theory, since the existing
approaches to two-phase systems, such as the classical
Maxwell-Garnet \cite{Maxwell1873,Maxwell1904} and Bruggeman
\cite{Bruggeman1935,Landauer1952}  mixing rules, their numerous
modifications (see
\cite{Bohren1983,Bergman1992,Sihvola1999,Tsang2001,Torquato2013,Milton2004}),
cluster expansions \cite{Torquato1984, Torquato1985b} for
dispersions of spheres with arbitrary degree of impenetrability,
their extensions \cite{Torquato1985} with the Pad\'e approximant
technique, and systematic simulations
\cite{MyroshnychenkoPRE2005,MyroshnychenkoJAP2005,Myroshnychenko2008,Myroshnychenko2010}
of random 2D systems of hard-core--penetrable-shell discs by
combining Monte Carlo algorithms and finite element calculations
do not exhibit it. The reason is that two-phase models
oversimplify the actual microstructure of CSEs and disregard the
processes involved.

A typical way out is to model a CSE as a three-phase system and
determine $\sigma_{\rm {eff}}$ by solving a pertinent
homogenization problem. The solution is expressed in terms of the
geometric and electric parameters of the phases. These parameters
are estimated so as to incorporate the relevant physical effects
and account for the observed behavior of $\sigma_{\rm eff}$.
Several classes of such models have been proposed.

(i) Cubic lattices of cubic insulators surrounded by highly
conductive layers and embedded in a conductive material
\cite{Wang1979,Jiang1995a,Jiang1995b}. The arrangement of
particles on a simple cubic lattice makes it possible to represent
the system with a resistor network and to calculate $\sigma_{\rm
eff}$ for the entire range of $c$. The results suggest that the
conductivity in the layer outside each particle may have a maximum
at certain distance away from the surface.

(ii) Three-component resistor models with a matrix represented by
normally conducting bonds, the inert randomly distributed
(quadratic or cubic) particles by insulating bonds, and the
interface region by highly conducting bonds
\cite{Bunde1985,Roman1986,Blender1987}. The models are solved by
Monte Carlo simulations or a position-space reorganization
technique and exhibit two threshold concentrations of the
insulating material. One corresponds to the onset of interface
percolation and the second one to a conductor-insulator
transition.  The effective medium and continuum percolation
approaches to these models are discussed in \cite{Rojo1988} and
\cite{Roman1987,Roman1990}, respectively.

\begin{figure}[bth]
\centering
\includegraphics[width=75mm]{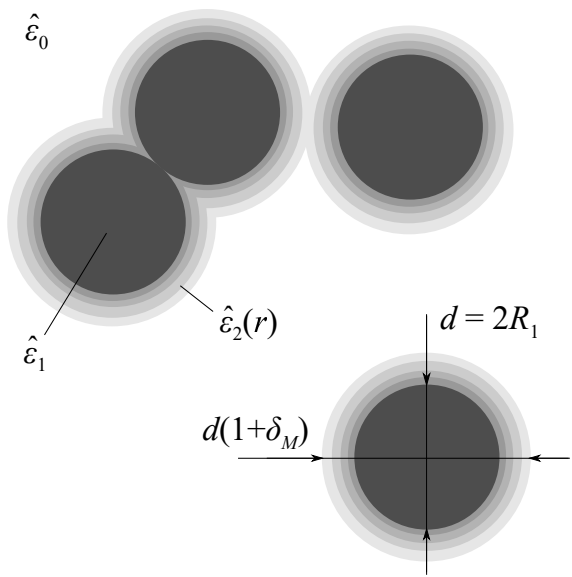}
\caption{\label{fig:model} The model under consideration. Each
dispersed particle consists of a spherical hard  core (shown
black) with diameter $d=2R_1$  and complex permittivity
$\hat{\varepsilon}_1$. The core is coated with an isotropic
(generally inhomogeneous) penetrable  concentric shell (graded
gray) with outer diameter $D= d(1+\delta_M)$, relative thickness
$\delta_M= (D-d)/d$, and complex permittivity profile
$\hat{\varepsilon}_2= \hat{\varepsilon}_2(r)$. The particles are
embedded in a uniform matrix (white) with complex permittivity
$\hat{\varepsilon}_0$. All the permittivities have form
(\ref{eq:ComplexPermittivity}). The local value of the
permittivity is determined by the shortest distance from the point
of interest to the nearest particle. }
\end{figure}

(iii) Random three-phase dispersions of spherical particles
comprising hard cores coated with concentric shells, either hard
or penetrable (see Fig.~\ref{fig:model}), of potentially higher
conductivity. Such core-shell models better suit the physical
conditions in CSEs, but are harder to analyze. Analytical studies,
such as
\cite{Stoneham1979,Brailsford1986,Nan1991L,Nan1991,Nan1993,Wiec1994},
of them are usually limited to the case of hard shells, involve a
sequence of one-particle approximations, and repeatedly use the
Maxwell-Garnet \cite{Maxwell1873,Maxwell1904} or/and Bruggeman
\cite{Bruggeman1935,Landauer1952} mixing rules. The case of
penetrable shells has been attacked through RRN simulations, such
as \cite{Siekierski2005,Siekierski2006,Siekierski2007} for mono-
and \cite{Kalnaus2011} for polysized particles. The essential
details of these core-shell model results are scrutinized in
Sec.~\ref{sec:simulations}.

In what follows, we derive a self-consistent analytic
many-particle solution for $\sigma_{\rm eff}$ of macroscopically
homogeneous and isotropic 3D model dispersions of
hard-core--penetrable-shell spheres, the shells being, in the
general case, electrically inhomogeneous and characterized by
radially-symmetrical, piecewise-continuous conductivity profiles
(see Fig.~\ref{fig:model} for the details of the model). The
desired $\sigma_{\rm eff}$ is a functional of the constituents'
conductivities and volume concentrations that satisfies a certain
integral relation, rigorous in the static limit. The volume
concentrations account for the statistical microstructure of the
system.

The derivation is carried out using the compact groups approach
(CGA)~\cite{Sushko2007,Sushko2009CompGroups,Sushko2009AnisPart,Sushko2017}.
It was originally designed to efficiently take into account
many-particle polarization and correlation effects, without an
in-depth modeling of those, in concentrated dielectric
dispersions. In this paper, elaborating its statistical-averaging
version, we (1) generalize the CGA to conducting systems whose
constituents have complex permittivities with first-order poles at
frequency $\omega =0$; (2)~scrutinize, for such systems, the
passage to the (quasi)static limit $\omega \to 0$ in all terms of
the iterative series for the averaged electric field and current;
(3) bring new arguments, not restricted to dielectric systems, for
the internally-consistent closure of the homogenization procedure
and determination of the complex permittivity of the auxiliary
host  matrix; and (4) propose a technique for dealing with
inhomogeneous overlapping regions.

Using the well-tested statistical physics results
\cite{Rikvold85,Lee1988,Rottereau2003} for the shell volume
concentration, we then validate the solution obtained by mapping
it onto the entire set of available 3D random resistor network
simulation data
\cite{Siekierski2005,Siekierski2006,Siekierski2007} for
dispersions of hard spheres coated with fully penetrable
(electrically uniform or inhomogeneous) concentric shells. The
solution is capable of recovering all of these data in the entire
ranges of $c$ simulated. To our best knowledge, no such an
analytic solution has been offered so far even for the simplest
case of uniform shells with their conductivity being equal to that
of the cores.

Finally, we apply the model to real CSEs. The results of
processing experimental data~\cite{Liang1973} clearly indicate
that the concept of inhomogeneous penetrable shells provides an
efficient way for describing the net effect on $\sigma_{\rm eff}$
by different mechanisms.  The latter may contribute most
significantly in different ranges of $c$. If so, they are
accounted for by different parts in the model shell conductivity
profile. This fact opens new opportunities for scrutinizing the
physics of processes in real composites, which is of crucial
importance in the situation where the consensus of opinions
regarding the nature of ionic conduction enhancement in various
composites has not been reached as yet
\cite{Agrawal1999,Uvarov2011,Kharton2011,Gao2016}.

The paper is arranged as follows. Some basic equations and
definitions of macroscopic electrodynamics for media with complex
permittivities of the constituents are recalled in
Sec.~\ref{sec:basics}. With those in mind, the CGA is generalized
in Sec.~\ref{sec:basicsCGA} to the problem of the effective
quasistatic complex permittivity $\hat{\varepsilon}_{\rm {eff}}$
of macroscopically homogeneous and isotropic dispersions. The
governing equation for $\hat{\varepsilon}_{\rm {eff}}$ is
expressed in terms of the statistical moments $\langle \left(
\delta \hat{\varepsilon} ({\rm {\bf r}}) \right)^{s} \rangle$ for
the local deviations of the permittivity distribution in the
dispersion from the complex permittivity $\hat{\varepsilon}_{\rm
{f}}$ of the host in the auxiliary system. By requiring that the
CGA and boundary conditions \cite{Sillars1937} for complex
electric fields be compatible, $\hat{\varepsilon}_{\rm {f}}$ is
determined in Sec.~\ref{sec:matrix}. The calculations of $\langle
\left( \delta \hat{\varepsilon} ({\rm {\bf r}}) \right)^{s}
\rangle$ for dispersions of isotropic core-shell particles with
electrically homogeneous and inhomogeneous shells are performed in
Secs.~\ref{sec:statapproach} and \ref{sec:inhomshells},
respectively. The resultant equations for $\sigma_{\rm eff}$ are
presented in Sec.~\ref{sec:conductivity}. Their validity is shown
in Sec.~\ref{sec:simulations} by mapping their solutions onto
extensive RRN simulation
data~\cite{Siekierski2005,Siekierski2007,Siekierski2006}. The
applicability of the theory to real ${\rm LiI/Al_2O_3}$
CSEs~\cite{Liang1973} is discussed in Sec.~\ref{sec:experiment}.
The main results of the paper  are summarized in
Sec.~\ref{sec:conclusion}.

\section{\label{sec:basics} Basic equations and definitions}

Consider the electromagnetic field caused in a nonmagnetic
heterogeneous medium by time-harmonic ($\sim e^{-i\omega t}$, $i$
being the imaginary unit) probing radiation whose working
frequencies $\omega$ are sufficiently small to neglect any
dielectric relaxation phenomena. The relevant frequency-domain
Maxwell's macroscopic equations, written in the Gaussian units,
have the form
\begin{equation} {\label{eq:MEdiv}}
{\rm div}\,{\bf D} =4 \pi \rho,\quad {\rm div}\,{\bf H} =0,
\end{equation}
\begin{equation} {\label{eq:MEcurl}}
{\rm curl}\,{\bf E} = i \frac{\omega}{c}\, {\bf H}, \quad {\rm
curl}\,{\bf H} = \frac{4\pi}{c}\,{\bf j}-i \frac{\omega}{c}\, {\bf
D},
\end{equation}
where ${\bf E}$, ${\bf D}$, ${\bf H}$, $\rho$, and ${\bf j}$ are
the amplitude distributions of the electric field, electric
displacement, magnetic field, free charge density, and free
current density, respectively, and $c$ is the speed of light in
vacuum. The densities $\rho$ and ${\bf j}$ are related by the
continuity equation
\begin{equation} {\label{eq:chargeConservation}}
 -i \omega \rho  + {\rm div}\,{\bf j} = 0.
\end{equation}

Assuming the standard linear  constitutive equations
\begin{equation} {\label{eq:materialEquation}}
{\bf D} =\varepsilon {\bf E}, \quad  {\bf j} =\sigma {\bf E},
\end{equation}
where $\varepsilon=\varepsilon({\bf r})$ and $\sigma=\sigma({\bf
r})$ are the local  (real) dielectric constant and electrical
conductivity in the medium, one can introduce the quasistatic
complex permittivity
 \begin{equation}
{\label{eq:ComplexPermittivity}} {\hat\varepsilon} = \varepsilon +
i \frac{4\pi \sigma}{\omega}
\end{equation}
of the medium to obtain from
Eqs.~(\ref{eq:MEdiv})--(\ref{eq:materialEquation}) the equation
for the quasistatic electric field distribution
\begin{equation} \label{basicEquation}
\Delta {\bf E} + k^{2}_0 \hat{\varepsilon}\, {\bf E} - {\rm
grad}\, {\rm div} {\textbf E} = 0,
\end{equation}
where $k_{0} = \omega/c$ is the magnitude of the wave vector ${\bf
k}_0$ of the incident field in vacuum, and define the complex
current density
\begin{equation} {\label{eq:ComplexCurrent}}
{\bf J} = -i\frac{\omega}{4\pi} \,{\hat\varepsilon}\,{\bf E}
,\quad {\rm div}\,{\bf J} = 0.
\end{equation}
The first Eq.~(\ref{eq:ComplexCurrent}) reduces in the static
limit ($\omega\to 0$) to Ohm's law, given by the second
Eq.~(\ref{eq:materialEquation}).

\section{\label{sec:basicsCGA} Compact group approach to homogenization of conducting systems}

The main points of this approach in application to macroscopically
homogeneous and isotropic nonconducting systems are discussed in
detail in
\cite{Sushko2007,Sushko2009CompGroups,Sushko2009AnisPart,Sushko2017}.
Here, closely following the summary in \cite{Sushko2017}, we
outline a generalization of the approach to macroscopically
homogeneous and isotropic dispersions $\cal{D}$ comprising
conducting constituents, such as conducting dielectrics or
imperfectly insulating materials. In view of the Kramers-Kronig
relations in the linear response theory, we assume that for a
given $\cal{D}$, the complex permittivities of all the
constituents have structure (\ref{eq:ComplexPermittivity}), where
$\varepsilon$ and $\sigma$ are in general piecewise-continuous and
bounded real functions of spacial coordinates; and that its
$\hat{\varepsilon}_{\rm eff} = \varepsilon_{\rm eff} + i 4\pi
\sigma_{\rm eff}/\omega$ can be calculated based upon the
following suggestions:

(1)  $\cal{D}$  is equivalent, in its response to a
long-wavelength probing field ($\omega \to 0$), to an auxiliary
system $\cal{S}$ prepared by embedding the constituents (particles
and matrix) of $\cal{D}$ into a uniform host (perhaps, imagined)
$\cal{M}$ with some permittivity $\hat {\varepsilon}_{\rm f}$.

(2) $\cal{S}$ can be viewed as a set of compact groups of both
particles and regions occupied by the real matrix. The compact
groups are defined as macroscopic regions whose typical sizes $d$
are much smaller than the wavelength $\lambda$ of probing
radiation in $\cal{M}$, but which yet include sufficiently large
numbers $N$ of particles to remain macroscopic and retain the
properties of the entire $\cal{S}$.

(3) The complex permittivity distribution in $\cal{S}$ is
\begin{equation} \label{profile}
\hat{\varepsilon}({\bf{r}}) = \hat{\varepsilon}_{\rm f} + \delta
\hat{\varepsilon}({\bf{r}})
\end{equation}
where $\delta\hat{\varepsilon} (\bf{r})$ is the contribution from
a compact group located at point $\bf{r}$. The explicit form of
$\delta{\hat\varepsilon} ({\bf{r}})$ is modeled according to the
geometrical and electrical parameters of $\cal{D}$'s constituents.

(4) $\hat{\varepsilon}_{\rm{eff}}$ can be found as the
proportionality coefficient in the relation
\begin{equation}
\label{eq:effcomplex} \langle {\bf{J}} ({\bf{r}})\rangle =
-i\frac{\omega}{4\pi}\langle \hat{\varepsilon} ({\bf{r}}) {\bf{E}}
({\bf{r}}) \rangle = -i\frac{\omega}{4\pi} \hat{\varepsilon}_{\rm
eff} \langle {\bf{E}} ({\bf{r}}) \rangle,
\end{equation}
where ${\bf{J}} ({\bf{r}})$ and ${\bf{E}} ({\bf{r}})$ are the
local values of the complex current and electric field,
respectively, and the angle brackets stand for the ensemble
averaging. In the static limit, provided $ \lim_{\omega \to
0}\omega {\varepsilon} ({\bf{r}})=0$ and ${\bf{E}} ({\bf{r}})$ is
real-valued, Eq.~(\ref{eq:effcomplex}) reduces to the common
definition \cite{Bergman1992,Sihvola1999,Torquato2013} of the
effective conductivity:
\begin{equation}
\label{eq:standardEeffconductivity} \langle \bf{j}
({\bf{r}})\rangle = \langle \sigma ({\bf{r}}) {\bf{E}} ({\bf{r}})
\rangle = \sigma_{\rm eff} \langle {\bf{E}} ({\bf{r}}) \rangle.
\end{equation}

(5) The electric field distribution $\textbf{E}(\textbf{r})$ in
$\cal{S}$ obeys the equation
\begin{equation} \label{a3}
\Delta {\bf E} + k_0^{2}\hat{\varepsilon}_{\rm f} {\bf E} - {\rm
grad}\, {\rm div} {\textbf E} = - k_{0}^{2} \delta
\hat{\varepsilon} {\bf E},
\end{equation}
which directly follows from Eqs.~(\ref{basicEquation}) and
(\ref{profile}).  The equivalent integral equation is
\begin{equation}
\label{eq4} {\rm {\bf E}}({\rm {\bf r}}) = {\rm {\bf E}}_{0} ({\rm
{\bf r}}) - {\int\limits_{V} {d{\rm {\bf {r}'}}\,}} {\rm T}(|{\rm {\bf
r}}-{\rm {\bf {r}'}}|) k_{0}^{2} \delta \hat{\varepsilon} ({\rm
{\bf {r}'}})\,{\rm {\bf E}}({\rm {\bf {r}'}}),
\end{equation}
where ${\rm {\bf E}}_{0} ({\rm {\bf r}}) = {\rm {\bf E}}_{0} \exp
(i\,{\rm {\bf k}} \cdot {\rm {\bf r}})$, ${\rm \bf{E}_{0}}$, and
${\bf k}= {\hat{\varepsilon} _{\rm f}}^{1/2} {\rm {\bf k}_0}$
(with ${\rm{Im}}\,{\hat{\varepsilon} _{\rm f}}^{1/2} \geq 0$) are,
respectively, the incident wave field, its amplitude, and its wave
vector in ${\cal M}$,  and ${\rm T}(r)$ is the Green's tensor of
Eq.~(\ref{a3}).

(6) The formal solution for ${\bf{E}}({\bf{r}})$ and those for
${\bf{J}}({\bf{r}})$, $\langle{\bf{E}}({\bf{r}})\rangle$, and
$\langle{\bf{J}}({\bf{r}})\rangle$ are representable in the form
of infinite iterative series. For systems whose constituents have
the permittivities of form (\ref{eq:ComplexPermittivity}), the
functions $k_0^2 \delta \hat{\varepsilon} ({\rm {\bf r}})$ in the
integrands and also the function $\omega \hat{\varepsilon}({\rm
{\bf r}})$ remain bounded even at $\omega\to 0$, where $|{\rm {\bf
k}}| \to 0$ as well. Mathematically, the situation is identical to
that for nonconducting systems and can be treated analogously.
Namely, in the iterative series for $\langle{\bf{E}}
({\bf{r}})\rangle$ and $\langle{\bf{J}}({\bf{r}})\rangle$, each
${\rm T}$ under the integral sign is replaced by its decomposition
${\widetilde {\rm T}}$, derived for a spherical exclusion volume
of radius $a\to 0$, into a Dirac delta function singular part and
a principal value part~\cite{Ryzhov1965,Weiglhofer1989}:
\begin{eqnarray}
\label{representation} {\widetilde T}_{\alpha\beta} ({\rm {\bf
r}}) = \frac{1}{3k^{2}} \delta_{\alpha\beta} \delta ({\rm {\bf
r}})\,e^{ikr} + \frac{1}{4\pi k^{2}}
\left(\frac{1}{r^3}-\frac{ik}{r^2}\right)\nonumber\\
\times\left( \delta _{\alpha\beta} - 3e_{\alpha} e_{\beta}
\right)\,e^{ikr} - \frac{1}{4\pi r}\left( {\delta _{\alpha\beta} -
e_{\alpha} e_{\beta}} \right)\,e^{ikr},
\end{eqnarray}
where $\delta ({\bf r})$ is the Dirac delta function,
$\delta_{\alpha\beta}$ is the Kronecker delta, and $e_{\alpha}$ is
the $\alpha$-component of the unit vector ${\bf e} = {\bf r}/r$.
The contributions to $\langle{\rm {\bf {E}} ({\bf r})}\rangle$
made by the subseries containing, in their integrands, the
principal value parts are estimated to be of the order
$|\hat{\varepsilon}_{\rm f}| k_0^2 L^3/d$, at most
\cite{Sushko2007}, as compared to those made by the subseries with
only the Dirac delta function parts. For a finite typical linear
size $L$ of the system, the former can be decreased below any
preset value by taking a sufficiently small $k_0$. So, passing to
the limit $\omega \to 0$ and formally replacing each ${\widetilde
{\rm T}}$  in the integrals for $\langle{\bf{E}}
({\bf{r}})\rangle$ and $\langle{\bf{J}}({\bf{r}})\rangle$ by its
Delta function part, we obtain
\begin{equation}\label{eq:E_averageZeroOmega}
\lim_{\omega \to 0}\left\langle
{\mathbf{E}}({\mathbf{r}})\right\rangle  = \lim_{\omega \to 0}
\left[ 1 + \langle \hat{Q} ({\mathbf{r}})\rangle
\right]{{\mathbf{E}}_0},
\end{equation}
\begin{equation}\label{eq:D_averageZeroOmega}
\lim_{\omega \to 0}
\left\langle{\mathbf{J}}({\mathbf{r}})\right\rangle    = - i
\lim_{\omega \to 0}
\frac{\omega\hat{\varepsilon}_{\rm{f}}}{4\pi}\left[ 1 - 2 \langle
\hat{Q} ({\mathbf{r}})\rangle \right]{{\mathbf{E}}_0},
\end{equation}
where
\begin{equation}\label{eq:sumZeroOmega}
\hat{Q}({\mathbf{r}})\equiv  \sum\limits_{s = 1}^\infty \left( -
\frac{1}{3\hat{\varepsilon} _{\rm{f}}} \right)^s (\delta
\hat{\varepsilon} ({\mathbf{r}}))^s.
\end{equation}

(7) The above results can be obtained without resort to iterative
series. Indeed, it follows from Eq.~(\ref{representation}) that
\begin{equation} \label{representationZerofrequency}
\lim_{\omega \to 0} k_{0}^2 \hat{\varepsilon}_{\rm f} {\widetilde
T}_{\alpha\beta} ({\rm {\bf r}}) = \tau^{(1)}_{\alpha\beta}+
\tau^{(2)}_{\alpha\beta},
\end{equation}
where
\begin{equation} \label{representationZerofrequencyParts}
\tau^{(1)}_{\alpha\beta}= \frac{1}{3}\,\delta _{\alpha\beta}
\delta ({\rm {\bf r}}), \quad \tau^{(2)}_{\alpha\beta} = \frac{
{\delta _{\alpha\beta} - 3e_{\alpha} e_{\beta}}}{4\pi r^3}.
\end{equation}
Substituting Eqs.~(\ref{representationZerofrequency}) and
(\ref{representationZerofrequencyParts}) into Eq.~(\ref{eq4}),
making simple algebraic manipulations and statistical averaging,
and implying that $\omega \to 0$, we obtain
\begin{equation}\label{eq:Eaveraged}
\begin{split}
\langle \mathbf{E} & (\mathbf{r}) \rangle = \left\langle \frac{ 3
\hat{\varepsilon}_{\rm f}  }{3 \hat{\varepsilon}_{\rm f}
+ \delta\hat{\varepsilon}({\bf r})} \right\rangle \mathbf{E}_0 \\
&- 3 \int\limits_V {d{\mathbf{r}}'} { \rm{\tau}^{(2)} (| {\bf r}
- {\bf r}' |) \left\langle \frac{\delta \hat{\varepsilon}
({\mathbf{r}}')}{3 \hat{\varepsilon}_{\rm f} + \delta
\hat{\varepsilon}({\bf r})}{\mathbf{E}}({\mathbf{r}}')
\right\rangle },
\end{split}
\end{equation}
\begin{equation}\label{eq:Caveraged}
\begin{split}
\langle \mathbf{J} & (\mathbf{r}) \rangle = -i \frac{\omega}{4\pi}
\, \hat{\varepsilon}_{\rm f}\left[1+ 2\left\langle \frac{
\delta\hat{\varepsilon}({\bf r}) }{3 \hat{\varepsilon}_{\rm f}
+ \delta\hat{\varepsilon}({\bf r})} \right\rangle \right] \mathbf{E}_0 \\
&+i \frac{3}{4\pi}\,  \int\limits_V {d{\mathbf{r}}'} {
\rm{\tau}^{(2)} (| {\bf r} - {\bf r}' |) \left\langle
\frac{\omega\, \hat{\varepsilon}({\bf r}) \,\delta
\hat{\varepsilon} ({\mathbf{r}}')}{3 \hat{\varepsilon}_{\rm f} +
\delta \hat{\varepsilon}({\bf r})}{\mathbf{E}}({\mathbf{r}}')
\right\rangle }.
\end{split}
\end{equation}
For  macroscopically isotropic and homogeneous systems, two-point
statistical averages depend only on $|{\bf r} - {\bf r}'|$. Due to
this symmetry and because of a special form of the angular
dependence of $\tau_{\alpha\beta}^{(2)}$, the integrals in
Eqs.~(\ref{eq:Eaveraged}) and (\ref{eq:Caveraged}) vanish.
Finally, viewing the expressions in the angle brackets as the sums
of infinite geometric series, we arrive at
Eqs.~(\ref{eq:E_averageZeroOmega})--(\ref{eq:sumZeroOmega}).

This consideration is very similar to that used in the
strong-property-fluctuation theory (SPFT)
\cite{Tsang2001,Ryzhov1965,Ryzhov1970,Tamoikin1971,Tsang1981,Zhuck1994,Michel1995,Mackay2000,Mackay2001}.
However, our theory gives another interpretation to $\delta
\hat{\varepsilon} ({\rm {\bf {r}}})$, appeals to the macroscopic
symmetry of the entire system instead of the symmetry of
correlation functions, and postulates no condition on the
stochastic field $ \hat{\xi} ({\bf r})= [\hat{\varepsilon} ({\bf
r})-\hat{\varepsilon}_{\rm f}]/[2\hat{\varepsilon}_{\rm
f}+\hat{\varepsilon} ({\bf r})]= \delta\hat{\varepsilon} ({\bf
r})/[3\hat{\varepsilon}_{\rm f}+\delta\hat{\varepsilon} ({\bf
r})]$ in order to improve the convergence of the iteration
procedure and decide on the value of $\hat{\varepsilon} _{\rm f}$.
Once the latter is determined, the analysis of $\hat{\varepsilon}
_{\rm eff}$ reduces to modeling $\delta \hat{\varepsilon} ({\rm
{\bf r}})$, calculating its moments $\langle \left( {\delta
\hat{\varepsilon} ({\rm {\bf r}})} \right)^{s} \rangle$, and
finding their sum in Eqs.~ (\ref{eq:E_averageZeroOmega}) and
(\ref{eq:D_averageZeroOmega}).

\section{\label{sec:matrix} Determination of $\hat{\varepsilon}_{\rm f}$}

If the permittivities of the constituents, that of $\cal{M}$, and
$\hat{\varepsilon}_{\rm eff}$  have at $\omega \to 0$ the
structure~({\ref{eq:ComplexPermittivity}), then, at least in this
limit, it is the Bruggeman-type of homogenization
$\hat{\varepsilon}_{\rm f}=\hat{\varepsilon}_{\rm eff}$ that is
compatible with the formalism of the CGA and definition
(\ref{eq:effcomplex}). To prove this statement, we first remind
that ${\bf E}_0$ is the amplitude of the probing electric field in
the uniform fictitious matrix of permittivity
$\hat{\varepsilon}_{\text f}$, and
$\left\langle{\mathbf{E}}({\mathbf{r}})\right\rangle $ is the
effective electric field in the homoge\-nized dispersion of
permittivity $\hat{\varepsilon}_{\rm eff}$, caused by the same
probing field. Next, we recall the boundary condition
\cite{Sillars1937}
$$\hat{\varepsilon}_1 E_{1n}=\hat{\varepsilon}_2 E_{2n} $$ for the
normal components of complex electric fields at the surface
between two conducting dielectrics (or imperfectly insulating
materials) with permittivities of type
(\ref{eq:ComplexPermittivity}). For the surface between the
fictitious matrix and the homoge\-nized dispersion, it gives
\begin{equation} \label{eq:boundarycondition}
\hat{\varepsilon}_{\rm
f} {\mathbf{E}}_{0n} = \hat{\varepsilon}_{\rm eff}
\left\langle{\mathbf{E}}({\mathbf{r}})\right\rangle_n.
\end{equation}
This relation, Eq.~(\ref{eq:effcomplex}), and
Eqs.~(\ref{eq:E_averageZeroOmega})--(\ref{eq:sumZeroOmega}) yield,
at $\omega \to 0$, the system of equations
\begin{equation}
\label{eq:system}
\begin{gathered}
\hat{ \varepsilon}_{\rm f} =\hat{\varepsilon}_{\rm eff}
\left(1+\langle \hat{Q}\rangle\right),\\
\hat{\varepsilon}_{\rm f}\left(1-2\langle \hat{Q}\rangle\right)
=\hat{\varepsilon}_{\rm eff} \left(1+\langle
\hat{Q}\rangle\right).
\end{gathered}
\end{equation}
Since $\hat{ \varepsilon}_{\rm f} \neq 0$, it follows immediately
that \begin{equation} \label{eq:ef} \hat{ \varepsilon}_{\rm f}
=\hat{ \varepsilon}_{\rm eff}
\end{equation}
and, for this $\hat{ \varepsilon}_{\rm f}$,
\begin{equation} \label{eq:equation}
\langle \hat{Q}({\mathbf{r}})\rangle=0.
\end{equation}
The latter is the desired governing equation for
$\hat{\varepsilon}_{\rm eff}$. It is valid in the limit $\omega
\to 0$.

It can be shown \cite{Sushko2017} that for two-constituent systems
(say, hard spheres embedded in a uniform host), Eqs.~(\ref{eq:ef})
and (\ref{eq:equation}) reproduce the Bruggeman
result~\cite{Bruggeman1935,Landauer1952,Ross2005} derived within
the assumption that spherical inclusions of all constituent
materials are placed in the effective medium. The same Eqs.
(\ref{eq:ef}), (\ref{eq:equation}), and Bruggeman result also
follow, in the quasistatic limit, from the SPFT \cite{Tsang1981}
for hard spheres where the condition $\langle \hat{\xi}({\bf r
})\rangle=0$ is imposed to eliminate the secular terms and the
bilocal approximation is implemented for a special case of the
spherically symmetric two-point correlation function
$\langle\hat{\xi}({\bf r})\hat{\xi}({\bf r'})\rangle$.

\section{\label{sec:statapproach} Statistical moments $\langle \left( \delta \hat{\varepsilon} ({\rm {\bf r}})
\right)^{s} \rangle$}

We consider a dispersion of $N$ spherically symmetrical core-shell
particles embedded into a uniform matrix. Suppose that the local
permittivity value at a point $\bf r$ within the dispersion is
determined by the distance $l\equiv\min\limits_{1\leq a\leq N}
|{\bf r}-{\bf r}_a |$ from $\bf r$  to the center of the nearest
ball as
\begin{equation} \hat{\varepsilon}({\bf r})=\begin{cases} \hat{\varepsilon}_1 &
{\text{if} }
\quad \,\,l<R_1,\\
\hat{\varepsilon}_2 & {\text{if} }\quad  R_1<l<R_2,\\
\hat{\varepsilon}_0 & {\text{if} }\quad \,\, l>R_2.
\end{cases} \label{distr}
\end{equation}
Here $R_1$ is the radius of the core of complex permittivity
$\hat{\varepsilon}_1$, $R_2$ is the outer radius of the shell of
complex permittivity $\hat{\varepsilon}_2$, and
$\hat{\varepsilon}_0$ is the complex permittivity of the matrix.
Within the CGA
\cite{Sushko2007,Sushko2009CompGroups,Sushko2009AnisPart,Sushko2017},
such a system can be modelled as follows.

Let $\theta(x)$ be the Heaviside step function and
$\chi_a^{(q)}({\bf r}) =\theta\left( R_q -|{\bf r}-{\bf
r}_a|\right)$ ($q= 0, 1, 2$) be the characteristic functions of
balls centered at point ${\bf r}_a$ and having radii $R_q$.
Suggesting that $R_1<R_2<R_0$ and allowing the balls to overlap,
consider the complex permittivity distribution of form
(\ref{profile}) with
\begin{eqnarray}
\delta\hat{\varepsilon} ({\bf r})&=& \Pi_1({\bf r})
\Delta\hat{\varepsilon}_1 +  \left[\Pi_2({\bf r})
-\Pi_1({\bf r}) \right] \Delta\hat{\varepsilon}_2 \nonumber\\
&&+  \left[\Pi_{0}({\bf r}) - \Pi_2({\bf r})\right]
\Delta\hat{\varepsilon}_0, \label{perm0}
\end{eqnarray}
where $\Delta \hat{\varepsilon}_q = \hat{\varepsilon}_q -
\hat{\varepsilon}_{\rm f}$, and each
\begin{equation}
\Pi_q({\bf r})=1-\prod\limits_{a=1}^N
\left(1-\chi_a^{(q)}\left({\bf r}\right)\right)
\label{ballsystemindicator}
\end{equation}
is the characteristic (indicator) function of the collection of
balls of radius $R_q$. In the limit $R_0 \to \infty $,
$\Pi_{0}({\bf r}) \to 1$ and Eq.~(\ref{perm0}) leads to the model
permittivity distribution ({\ref{distr}}) for a dispersion of
penetrable core-shell particles embedded into a uniform matrix of
permittivity $\hat{\varepsilon}_0$.  Note that $R_0$ is a
convenient auxiliary parameter (having nothing to do with the
physical geometry) which is used for the host matrix to be
introduced within the same formal algorithm as the other
constituents are. The characteristic function of the entire region
occupied by the substance of permittivity $\hat {\varepsilon_q}$
in this dispersion is given by the coefficient function in front
of the corresponding $\Delta \hat{\varepsilon}_q$. It is readily
verified that the characteristic functions of regions with
different permittivities are mutually orthogonal.

Further, we  limit ourselves to the case of particles with hard
cores. Then $\chi_{a}^{(1)}({\bf r})\chi_{b}^{(1)}({\bf
r})=\delta_{ab}\chi_{a}^{(1)}({\bf r}),$ where $\delta_{ab}$ is
the Kronecker delta, and $\Pi_1({\bf r})$ reduces to
\begin{equation}
\Pi_1({\bf r}) = \sum\limits_{a=1}^N \chi_a^{(1)}\left({\bf
r}\right), \label{pi1}
\end{equation}
with the additional restriction $|{\bf r}_a-{\bf r}_b|\geq 2R_1$
on the locations of any two balls.

For a macroscopically homogeneous and isotropic system
$$\left\langle\sum\limits_{a=1}^N
\chi_a^{(1)}\left({\bf r}\right)\right\rangle=c, $$ where $c$ is
the volume concentration of the hard cores. In view of this fact
and the mutual orthogonality of the characteristic functions of
regions with different permittivities, the moments of the function
(\ref{perm0})  can be represented in the limit $R_0 \to \infty $
as
\begin{eqnarray}
\langle\left[\delta\hat{\varepsilon}({\bf r})\right]^s\rangle &=&
c\left(\Delta\hat{\varepsilon}_1\right)^s
+\left(\phi(c,\delta)-c\right)\left(\Delta\hat{\varepsilon}_2\right)^s \nonumber\\
&&+\left(1-\phi(c,\delta)\right)\left(\Delta\hat{\varepsilon}_0\right)^s,
\label{perm2}
\end{eqnarray}
where
\begin{eqnarray}
\phi(c,\delta) &\equiv &  \left\langle\Pi_2({\bf r}) \right\rangle
= \left\langle\left[1-\prod\limits_{a=1}^N
\left(1-\chi_a^{(2)}\left({\bf
r}\right)\right)\right] \right\rangle \nonumber\\
&=&\left\langle \sum\limits_{1 \leq a \leq N}
{\chi}_a^{\,(2)}\left({\bf r}\right)-\sum\limits_{1 \leq a < b
\leq N} {\chi}_a^{\,(2)}\left({\bf
r}\right){\chi}_b^{\,(2)}\left({\bf r}\right) \right. \nonumber \\
&&+ \left.\sum\limits_{1 \leq a < b < c \leq N}
{\chi}_a^{\,(2)}\left({\bf r}\right){\chi}_b^{\,(2)}\left({\bf
r}\right){\chi}_c^{\,(2)}\left({\bf r}\right)-\ldots \right\rangle
\nonumber\\
\label{eq:effectiveconcentration}
\end{eqnarray}
is the effective volume concentration of
hard-core--penetrable-shell particles \cite{Torquato2013}. Besides
$c$, it depends on the relative thickness of the shell $\delta =
\left(R_2-R_1 \right)/R_1$; in particular, $\phi(c,0)=c$. The
averaged values of the sums in
Eq.~(\ref{eq:effectiveconcentration}) are calculated using the
partial distribution functions $F_s({\bf r}_1, {\bf r}_2,
\ldots,{\bf r}_s)$ for the system under consideration.

For hard-core--hard-shell particles
Eq.~(\ref{eq:effectiveconcentration}) gives
\begin{equation} \label{hard} \phi (c,\delta) =c (1+\delta)^3.
\end{equation}

\section{\label{sec:inhomshells} The Case of Inhomogeneous Isotropic Shells}

To extend the results of Sec.~\ref{sec:statapproach} to
dispersions of spherically symmetrical particles with hard cores
and adjacent inhomogeneous penetrable shells, we begin with the
situation where each shell consists of $M$ concentric spherical
layers with outer radii $R_{2,m}$ (grouped in the order of
increasing magnitude) and constant dielectric permittivities
$\hat{\varepsilon}_{2,m}$, $m= 1, 2,\dots, M$. Next, we suggest
that the local permittivity distribution within the dispersion is
given by this law, generalizing Eq. (\ref{distr}):
\begin{equation} \hat{\varepsilon}({\bf r})=\begin{cases}
\hat{\varepsilon}_1 & {\text{if} }
\quad \,\,l<R_1,\\
\hat{\varepsilon}_{2,1} & {\text{if} }\quad  R_1<l<R_{2,1},\\
\hat{\varepsilon}_{2,m} & {\text{if} }\quad  R_{2,m-1}<l<R_{2,m}, \,2 \leq m \leq M,\\
\hat{\varepsilon}_0 & {\text{if} }\quad \,\, l>R_{2,M}.
\end{cases} \label{distr1}
\end{equation}

Let  $\chi_a^{(2, m)}({\bf r}) =\theta\left( R_{2, m} -|{\bf
r}-{\bf r}_a|\right)$ be the characteristic functions of balls
centered at point ${\bf r}_a$, having radii $R_{2,m}$, and allowed
to overlap. Then the characteristic functions of the collections
of balls with radii $R_{2,m}$ are
\begin{equation} \Pi_{2,m}({\bf r})= 1-\prod \limits_{a=1}^N
\left(1-\chi_a^{(2, m)}\right).
\label{characteristicfunctionregions}
\end{equation}
Repeating almost literally the reasoning in
Sec.~\ref{sec:statapproach}, we can represent the distribution
(\ref{distr1}) in form (\ref{profile}) with
\begin{eqnarray}
\delta\hat{\varepsilon} ({\bf r})&=& \left[1- \Pi_{2,M}({\bf
r})\right] \Delta\hat{\varepsilon}_0
+ \Pi_1({\bf r}) \Delta\hat{\varepsilon}_1\nonumber\\
&+&  \left[\Pi_{2,1}({\bf r})
-\Pi_1({\bf r})  \right] \Delta\hat{\varepsilon}_{2,1}\nonumber\\
&+& \sum\limits_{m=2}^M  \left[ \Pi_{2,m}({\bf r}) -
\Pi_{2,m-1}({\bf r}) \right] \Delta\hat{\varepsilon}_{2,m},
\label{perm00}
\end{eqnarray}
where $\Delta \hat{\varepsilon}_{2,m} = \hat{\varepsilon}_{2,m} -
\hat{\varepsilon}_{\rm f}$. Correspondingly,
\begin{eqnarray}
\langle\left[\delta\hat{\varepsilon}({\bf r})\right]^s\rangle &=&
\left[1-\phi(c,\delta_M)\right]\left(\Delta\hat{\varepsilon}_0\right)^s + c\left(\Delta\hat{\varepsilon}_1\right)^s \nonumber \\
&+&\sum\limits_{m=1}^M\left[\phi(c,\delta_m)-\phi(c,\delta_{m-1})\right]\left(\Delta\hat{\varepsilon}_{2,m}\right)^s,
\label{perm22}
\end{eqnarray}
where $\delta_m =\left(R_{2,m}-R_1\right)/R_1 $, $\phi(c,\delta_m)
\equiv \left\langle \Pi_{2,m}({\bf r})\right\rangle$ is given by
Eq.~(\ref{eq:effectiveconcentration}) at $\delta=\delta_m$, and we
denoted $\delta_0=0$. Finally, passing to the limits $M \to
\infty$, $| \delta_m-\delta_{m-1}| \to 0$ ($\delta_M = {\rm
const}$) and assuming $\phi(c,\delta)$ to be differentiable with
respect to $\delta$, for a dispersion of particles with a
piecewise-continuous complex permittivity profile
$\hat{\varepsilon}_2 ({r})$ of the shells we obtain
\begin{eqnarray}
\langle\left[\delta\hat{\varepsilon}({\bf r})\right]^s\rangle
&=&\left[1-\phi(c,\delta_M)\right]\left(\Delta\hat{\varepsilon}_0\right)^s
+ c\left(\Delta\hat{\varepsilon}_1\right)^s \nonumber \\
&+&\int\limits_0^{\delta_M}\frac{\partial \phi(c,u)}{\partial
u}\left[\Delta\hat{\varepsilon}_2(u)\right]^s du,
\label{eq:nonuniformmoments}
\end{eqnarray} where $\Delta \hat{\varepsilon}_2(u)$ is the
deviation $\hat{\varepsilon}_2(r)-\hat{\varepsilon}_{\rm f}$ as a
function of $u=(r-R_1)/R_1$ and $\delta_M$ corresponds to the
outermost edge of the shell.

For a uniform shell ($\Delta \hat{\varepsilon}_2 = {\rm const}$),
Eq. (\ref{eq:nonuniformmoments}) immediately reduces to Eq.
(\ref{perm2}) with $\delta = \delta_M=\delta_1$.

\section{\label{sec:conductivity} Equations for Effective Conductivity}

In the case of uniform shells, where the moments
$\langle\left[\delta\hat{\varepsilon}({\bf r})\right]^s\rangle$
are given by Eq.~(\ref{perm2}), the sums involved in
Eq.~(\ref{eq:equation}) take the form $\sum_{s=1}^\infty \left(
-{\Delta \hat{\varepsilon}_q }/{3\hat{\varepsilon}_{\rm
eff}}\right)^s$. For $|{\Delta \hat{\varepsilon}_q
}/{3\hat{\varepsilon}_{\rm eff}}|<1$, they reduce to infinite
geometric series, so
\begin{equation} \label{eq:sum}
\sum\limits_{s=1}^\infty \left(-\frac{\Delta \hat{\varepsilon}_q
}{3\hat{\varepsilon}_{\rm eff}}\right)^s=
-\frac{\hat{\varepsilon}_q -\hat{\varepsilon}_{\rm
eff}}{2\hat{\varepsilon}_{\rm eff}+\hat{\varepsilon}_q}.
\end{equation}
For $|{\Delta \hat{\varepsilon}_q }/{3\hat{\varepsilon}_{\rm
eff}}|\geq1$, the left-hand side in Eq.~(\ref{eq:sum}) can be
treated, as was shown in Sec.~\ref{sec:basicsCGA}, as an
asymptotic series of the right-hand side, so that the restriction
$|{\Delta \hat{\varepsilon}_q }/{3\hat{\varepsilon}_{\rm eff}}|<1$
can be omitted. The resulting equation for $\hat{\varepsilon}_{\rm
eff}$ is
\begin{eqnarray}
\left[1-\phi(c, \delta)\right]\frac{\hat{\varepsilon}_0
-\hat{\varepsilon}_{\rm eff}}{2\hat{\varepsilon}_{\rm
eff}+\hat{\varepsilon}_0} + c\,\frac{\hat{\varepsilon}_1
-\hat{\varepsilon}_{\rm eff}}{2\hat{\varepsilon}_{\rm
eff}+\hat{\varepsilon}_1}
\nonumber\\
+\left[\phi(c, \delta)-c\right]\frac{\hat{\varepsilon}_2
-\hat{\varepsilon}_{\rm eff}}{2\hat{\varepsilon}_{\rm
eff}+\hat{\varepsilon}_2}=0. \label{eq:uniformshell}
\end{eqnarray}

To extract the equation for the quasistatic $\sigma_{\rm eff}$, we
pass in Eq.~(\ref{eq:uniformshell}) to the limit $\omega \to 0$
and assume that
\begin{equation} \label{eq:conditions}
\begin{gathered} 2\sigma_{\rm eff} +\sigma_q \gg  \frac{\omega}{4\pi}
\left(2\varepsilon_{\rm eff} +\varepsilon_q \right), \\
|\sigma_q -\sigma_{\rm eff}| \gg \frac{\omega}{4\pi} |
\varepsilon_q -\varepsilon_{\rm eff} |.
\end{gathered}
\end{equation}
Then, retaining the first term in the formal perturbation series
in $\omega$ for the left-hand side of Eq.~(\ref{eq:uniformshell}),
we obtain
\begin{eqnarray}
\left[1-\phi(c, \delta)\right]\frac{\sigma_0 -\sigma_{\rm
eff}}{2\sigma_{\rm eff}+\sigma_0} + c\,\frac{\sigma_1 -\sigma_{\rm
eff}}{2\sigma_{\rm eff}+\sigma_1} \nonumber \\
+\left[\phi(c, \delta)-c\right]\frac{\sigma_2 -\sigma_{\rm
eff}}{2\sigma_{\rm eff}+\sigma_2}=0. \label{eq:conductivity}
\end{eqnarray}

In view of Eq.~({\ref{eq:conditions}}), the sufficient condition
for the validity of Eq.~(\ref{eq:conductivity}) can be represented
as
\begin{equation} \label{eq:sufficentcondition}
|\sigma_{\rm eff}-\sigma_{ q}|  \gg  \frac{\omega}{4\pi}
\left(2\varepsilon_{\rm eff} +\varepsilon_q \right), \quad q =1,
2, 3.
\end{equation}

The generalizations of Eqs.~(\ref{eq:uniformshell}) and
(\ref{eq:conductivity}) to dispersions of particles with
inhomogeneous isotropic shells are evident [see
Eq.~(\ref{eq:nonuniformmoments})]:
\begin{eqnarray}
\left[1-\phi(c, \delta_M)\right]\frac{\hat{\varepsilon}_0
-\hat{\varepsilon}_{\rm eff}}{2\hat{\varepsilon} _{\rm
eff}+\hat{\varepsilon}_0}+ c\,\frac{\hat{\varepsilon}_1
-\hat{\varepsilon}_{\rm eff}}{2\hat{\varepsilon}_{\rm
eff}+\hat{\varepsilon}_1} \nonumber \\
+\int\limits_0^{\delta_M}\frac{\partial \phi(c,u)}{\partial
u}\frac{\hat{\varepsilon}_2 (u) -\hat{\varepsilon}_{\rm
eff}}{2\hat{\varepsilon}_{\rm eff}+\hat{\varepsilon}_2 (u)}\,du
=0,  \label{eq:nonuniformshell}
\end{eqnarray}
\begin{eqnarray}
\left[1-\phi(c, \delta_M)\right]\frac{\sigma_0 -\sigma_{\rm
eff}}{2\sigma _{\rm eff}+\sigma_0}+ c\,\frac{\sigma_1 -\sigma_{\rm
eff}}{2\sigma_{\rm eff}+\sigma_1} \nonumber
\\
+\int\limits_0^{\delta_M}\frac{\partial \phi(c,u)}{\partial
u}\frac{\sigma_2 (u) -\sigma_{\rm eff}}{2\sigma_{\rm eff}+\sigma_2
(u)}\,du =0.
\label{eq:nonuniformconductivity}
\end{eqnarray}

Based upon the volume averaging procedure,
Eqs.~(\ref{eq:uniformshell}) and (\ref{eq:conductivity}) were
first proposed in \cite{Sushko2013}, and
Eqs.~(\ref{eq:nonuniformshell}) and
(\ref{eq:nonuniformconductivity}) in \cite{Sushko2018polymer}.
Here,  Eqs.~(\ref{eq:conductivity}) and
(\ref{eq:nonuniformconductivity}) are finally substantiated with a
statistical mechanics formalism  and an internally closed
homogenization procedure.

Note that care must be taken when applying Eqs.
(\ref{eq:conductivity}) and (\ref{eq:nonuniformconductivity}) to
experimental data. In practice, $\sigma_{\rm eff}$ is often
identified with the quasistatic conductivity recovered from
impedance measurements at very small (say, $\omega/2\pi \leq
1\,{\rm kHz}$), yet nonzero frequencies. Equations
(\ref{eq:conductivity}) and (\ref{eq:nonuniformconductivity})
remain applicable to such situations as long as all inequalities
(\ref{eq:conditions}) hold true for the real and imaginary parts
of the quasistatic complex permittivities of the constituents.

We complete this section by mentioning that various mixing rules
of the Maxwell-Garnett type are formally obtainable within the CGA
by setting $\hat{\varepsilon}_{\rm f}=\hat{\varepsilon}_0$
\cite{Sushko2007,Sushko2009CompGroups,Sushko2009AnisPart,Sushko2017}.
Then the moments (\ref{eq:nonuniformmoments}) take the form
$$\langle\left[\delta\hat{\varepsilon}({\bf r})\right]^s\rangle =
c\left(\hat{\varepsilon}_1-\hat{\varepsilon}_0\right)^s
+\int\limits_0^{\delta_M}\frac{\partial \phi(c,u)}{\partial
u}\left[\hat{\varepsilon}_2(u)-\hat{\varepsilon}_0\right]^s du$$
to eventually give, in the quasistatic limit,
\begin{equation}
\sigma_{\rm eff} = \sigma_0\,\frac{1+2F}{1-F}
\label{eq:MGapproach}
\end{equation} where
$$F= c \frac{\sigma_1 -\sigma_0}{2 \sigma_0+\sigma_1}+\int\limits_0^{\delta_M}\frac{\partial \phi(c,u)}{\partial
u}\frac{\sigma_2(u)-\sigma_0}{2\sigma_0 + \sigma_2(u)}\, du.$$ For
the shells consisting of $M$ concentric spherical layers with
constant conductivities $\sigma_{2,m}$, the last integral equals
$$ \sum\limits_{m=1}^M\left[\phi(c,\delta_m)-\phi(c,\delta_{m-1})\right]
\frac{\sigma_{2,m}-\sigma_0}{2 \sigma_0+ \sigma_{2,m}}.$$

\section{\label{sec:simulations} Comparison with analytical results and numerical simulations}

Before  we proceed to processing experimental data with the above
theory, it is of crucial importance to test its validity by
contrasting it with other authors' analytical and computer
simulation results.

For 3D systems of particles with hard cores and fully penetrable
shells calculations \cite{Rikvold85}, done within the
scaled-particle approximation \cite{Reiss1959} for hard-sphere
fluids, give
\begin{eqnarray}\phi(c,\delta)= 1-
(1 - c)\,\exp\left[{-\frac{((1+\delta)^3 -
1)c}{1-c}}\right] \nonumber\\
 \times  \exp\left\{- \frac{3(1 + \delta)^3
c^2}{2(1 - c)^3} \left[2 - \frac{3}{1+\delta} +
\frac{1}{(1+\delta)^3} \right. \right. \nonumber \\
\label{effconc1} - \left. \left. \left( \frac{3}{1+\delta} -
\frac{6}{(1+\delta)^2} + \frac{3}{(1+\delta)^3}\right) c
\right]\right\}. \label{effectiveconcentration}
\end{eqnarray}
This result is in very good agreement with Monte Carlo simulations
\cite{Lee1988} (see also \cite{Rottereau2003}). Simulation results
for  $\sigma_{\rm eff}$ (and $\phi(c,\delta)$) of 3D systems of
monosized particles are available in
\cite{Siekierski2005,Siekierski2007,Siekierski2006}, where
$\sigma_{\rm eff}$ was calculated using the RRN approach
\cite{Kirkpatrick1971,Kirkpatrick1973}.

In simulations
\cite{Siekierski2005,Siekierski2007,Siekierski2006}, the virtual
RRN samples, representing the morphology and phase structure of
simulated dispersions, were built from a matrix with $300^3$ cubic
cells by placing spherical grains randomly into the matrix and
using a special algorithm for avoiding conflicts of spatial
restrictions on the grain locations. Each cell was marked as
belonging to a grain  if the center of the cell lay inside it. The
procedure was repeated until the assumed volume fraction $c$ of
the filler was attained. The residual part of the virtual sample
(not belonging to the grains) was attributed to either the shells,
with prescribed thickness $t$, or the matrix (all cells belonging
to neither the grains nor the shell). Finally, each cell was
represented as a parallel combination of a resistor and a
capacitor, with their parameters taken according to the assumed
material parameters of all the phases; the electrical parameters
used are summarized in Table~\ref{tab:simulationparameters}.
Replacing each pair of neighboring cells by an equivalent
electrical circuit with the corresponding impedance and the nodes
at the centers of the cells, the virtual samples were analyzed as
3D networks of such impedances.

We use results
\cite{Rikvold85,Siekierski2005,Siekierski2007,Siekierski2006} to
test the validity of Eqs.~(\ref{eq:conductivity}) and
(\ref{eq:nonuniformconductivity}) in the following five steps.

\begin{table}[tbh]
\caption{\label{tab:simulationparameters} Electrical parameters,
in ${\rm{S/cm}}$, used in simulations
\cite{Siekierski2005,Siekierski2007} (shells with constant
conductivity) and \cite{Siekierski2006} (shells with inhomogeneous
conductivity profiles (\ref{eq:GaussianSimulations})).}
\begin{ruledtabular}
\begin{tabular}{lccccc}
Simulations  & $\sigma_0$ & $\sigma_1$ & $\sigma_2$ & $\sigma_{\rm min}^\prime$ & $\sigma_{\rm max}^\prime$ \\
\hline
\cite{Siekierski2005,Siekierski2007}  & $1\times 10^{-8}$ & $1\times 10^{-12}$ & $1\times 10^{-4}$     & \\
\cite{Siekierski2006}  & $1\times 10^{-8}$ & $1\times 10^{-12}$ &   & $1\times 10^{-6}$ & $1\times 10^{-4}$\\
\end{tabular}
\end{ruledtabular}
\end{table}

\subsection{\label{step1} Mapping the geometrical parameters in
models~\cite{Rikvold85,Siekierski2007} onto each other}

This step is to contrast results \cite{Rikvold85,Siekierski2007}
for $\phi(c,\delta)$ in order to find the relations between the
parameters $c$ and $\delta$  for a dispersion of spherical
core-shell particles \cite{Rikvold85} and their counterparts
$c^\prime$ and $\delta^\prime$ in simulations
\cite{Siekierski2005,Siekierski2007,Siekierski2006}.

Evidently, for a given $t$ and under the condition that the
assumed volume fraction $c$ of the filler is attained,
$c^\prime=c$, the simulation procedure \cite{Siekierski2007} leads
to values of $\delta^\prime$ different from those of $\delta$.
Indeed, suppose that $N$ spherical grains of radius $a/2$, where
$a$ is the edge of the cubic cell, are used to generate a virtual
sample of volume $V$. Then $c^\prime = Na^3/V$ and $\delta^\prime
=2t/a$, for only one cell can belong to each grain. To achieve
this filler concentration in the same-volume dispersion of $N$
spherical particles with shell thickness $t$, their core radius
must be equal to $(4\pi/3)^{-1/3}a$. For such particles, the
relative shell thickness $\delta =(4\pi/3)^{1/3} t/a$.
Correspondingly,
\begin{equation} \delta = K
\delta^\prime, \label{eq:mapping}
\end{equation}
where, in our example, $K=k\equiv(\pi/6)^{1/3}\approx 0.806$.
Considering $K$ as a fitting parameter, one can generalize
Eq.~(\ref{eq:mapping}) to the situations where each grain contains
a large number of cells. The greater this number is,  the closer
to unity  $K$ is expected to be.  However, for $a = 0.5\,{\rm \mu
m}$, used in \cite{Siekierski2005,Siekierski2007,Siekierski2006},
the deviation of $K$ from unity should be noticeable.

The results of applying Eqs.~(\ref{effectiveconcentration}) and
(\ref{eq:mapping}) to simulation data \cite{Siekierski2007} for
the composition of dispersions of spherical
hard-core--penetrable-shell particles  at different values of $c$
are presented in Figs.~\ref{fig:SiekierskiShell_107} and
\ref{fig:SiekierskiShell_103-9}. They clearly demonstrate that
under the proper choice of $K$, these equations describe data
\cite{Siekierski2007} very well; the found values of $K$ turn out
to be close to our above estimates.

\begin{figure}
\centering
\includegraphics[width=75mm]{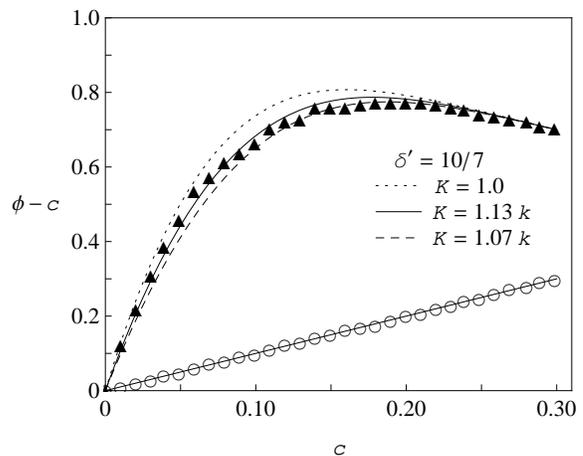}
\caption{ \label{fig:SiekierskiShell_107} Simulation data
\cite{Siekierski2007} for the volume concentration ($\phi-c$,
$\blacktriangle$) of freely-penetrable shells of thickness $t =
5\,{\rm \mu m}$ as a function of the assumed volume concentration
($c$, $\circ$) of hard grains of diameter $d= 7\,{\rm \mu m}$, and
the fits to these data with Eq.~(\ref{effectiveconcentration}) for
different values of the mapping parameter $K$ in
Eq.~(\ref{eq:mapping}).}
\end{figure}

\begin{figure}
\centering
\includegraphics[width=75mm]{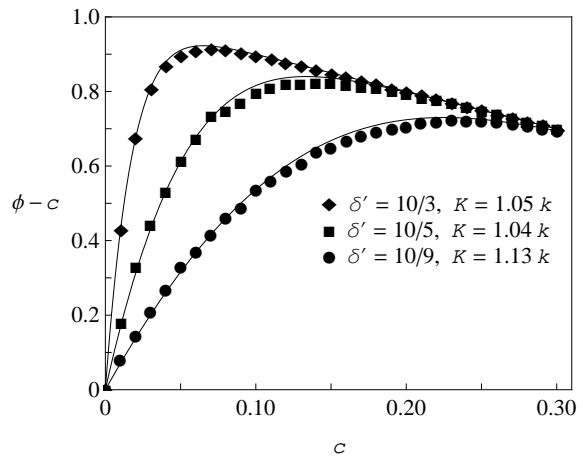}
\caption{\label{fig:SiekierskiShell_103-9} Simulation data
\cite{Siekierski2007} and the fits analogous to those in
Fig~\ref{fig:SiekierskiShell_107}, but for $d = 3$
($\blacklozenge$), 5 ($\blacksquare$), and $9\,{\rm \mu m}$
($\bullet$) at fixed $t = 5\,{\rm \mu m}$. }
\end{figure}

\subsection{\label{step2} Verifying functional
relationship~(\ref{eq:conductivity}) between $\sigma_{\rm eff}$
and $\phi$}

With this object in view, the dispersion composition is assumed to
be known for different values of $c$ at fixed $t$ and $d$. Taking
the corresponding values  of $\phi$ from simulations
\cite{Siekierski2007} (Figs.~\ref{fig:SiekierskiShell_107} and
\ref{fig:SiekierskiShell_103-9}), we then use
Eq.~(\ref{eq:conductivity}) to calculate $\sigma_{\rm eff}$ as a
function of $c$ for given $t$ and $d$ without referring to
Eqs.~(\ref{effectiveconcentration}) and (\ref{eq:mapping}).

The results so obtained are shown in
Fig.~\ref{fig:SiekierskiShellConductivity_105-9}, together with
conductivity simulation results \cite{Siekierski2007}.  If $c
\gtrsim 0.07$, the agreement between both theories is good for all
three sets of data ($d = 5$, 7, and $9\,{\rm \mu m}$ at fixed $t =
5\,{\rm \mu m}$). At lower values of $c$, our theory predicts the
percolation-type behavior of $\sigma_{\rm eff}$ (see also
Fig.~\ref{fig:SiekierskiConductivity107a_4} and the inset), with
the threshold concentration $c_{\rm c}$ that can be estimated from
the relation $\phi =1/3$ \cite{Sushko2013}. For the indicated sets
of data, the estimations with Eqs.~(\ref{eq:conductivity}) and
(\ref{eq:mapping}) give, respectively, $c_{\rm c} = 0.020$ ($K/k =
1.04$, $d=5\,{\rm \mu m}$), $0.034$ ($K/k = 1.07$, $d=7\,{\rm \mu
m}$), and 0.046 ($K/k = 1.13$, $d=9\,{\rm \mu m}$). Contrastingly,
the simulated values of $\sigma_{\rm eff}$ seem to increase
gradually even at the  lowest values of $c$, and the percolation
thresholds, if any, are hard to detect. This situation is typical
of conductivity simulations for finite-size systems where the
percolation threshold is a random non-Gaussian variable
\cite{Berlyand1997}.

\begin{figure}
\centering
\includegraphics[width=75mm]{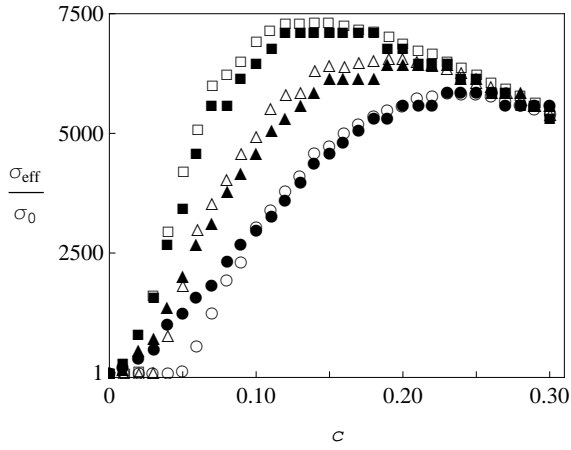}
\caption{\label{fig:SiekierskiShellConductivity_105-9} Simulation
data \cite{Siekierski2007} for $\sigma_{\rm eff}$ as a function of
$c$ for a fixed shell thickness $t= 5\,{\rm \mu m}$ and grain
diameters $d = 5$ ($\blacksquare$), 7 ($\blacktriangle$), and
$9\,{\rm \mu m}$ ($\bullet$). Empty symbols ($\square$,
$\triangle$, and $\circ$): our corresponding results obtained from
Eq.~(\ref{eq:conductivity}) by setting $\phi$ equal to the
simulated values \cite{Siekierski2007}, shown in
Figs.~\ref{fig:SiekierskiShell_107} and
\ref{fig:SiekierskiShell_103-9}.}
\end{figure}

\begin{figure}
\centering
\includegraphics[width=75mm]{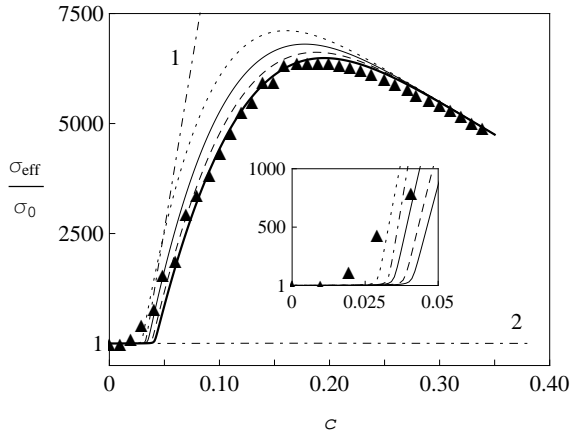}
\caption{\label{fig:SiekierskiConductivity107a_4} Simulation data
\cite{Siekierski2007} for $\sigma_{\rm eff}$ as a function of $c$
at  $t= 5\,{\rm \mu m}$ and  $d = 7\,{\rm \mu m}$
($\blacktriangle$), and their fits by Eq.~(\ref{eq:conductivity})
with $\phi(c,\delta)$ given by Eq.~(\ref{effectiveconcentration})
for the values of $\delta$ from Fig.~\ref{fig:SiekierskiShell_107}
(dotted, dashed, and solid lines) and $K=1.03\, k$ (thick solid
line). The dotdashed lines: the result for  $\sigma_{\rm eff}$
given by Eq.~(\ref{eq:conductivity}) with $\phi(c,\delta)$ for
hard shells [see Eq.~(\ref{hard})] and $K=1.07 k$  (line 1); that
by the Maxwell-Garnet-type Eq.~(\ref{eq:MGapproach}) with
$\phi(c,\delta)$ given by Eq.~(\ref{effectiveconcentration}) and
$K=1.03 k$ (line 2). Note that: both results fail to reproduce
data \cite{Siekierski2007}; line 2 has a maximum of $x_{\rm eff }
\approx 7.2$ at $c\approx 0.17$ which cannot be resolved in this
figure.}
\end{figure}

\subsection{\label{step3} Testing our model for the case of uniform penetrable
shells}

 This step consists in fitting conductivity data
\cite{Siekierski2005,Siekierski2007} using
Eq.~(\ref{eq:conductivity}) with $\phi =\phi(c,\delta)$ given by
Eq.~(\ref{effectiveconcentration}) and $\delta$ given by
Eq.~(\ref{eq:mapping}). As
Fig.~\ref{fig:SiekierskiConductivity107a_4} demonstrates, the
value of $K\approx 1.07 k$, determined by fitting the composition
data \cite{Siekierski2007} (step~A), is also appropriate to
reproduce conductivity data \cite{Siekierski2007} (this step).
Similarly, Figs.~\ref{fig:Siekierski_HomogeneousLayers_t_fixed}
and \ref{fig:Siekierski_HomogeneousLayers_d_fixed} clearly
indicate that the parameter $K$ alone, with a reasonable fitting
value for each series, is sufficient to reproduce all ten series
of simulation data \cite{Siekierski2007} for $\sigma_{\rm eff}$ of
dispersions of particles with uniform penetrable shells. This fact
is a strong argument in favor of the model expressed by
Eqs.~(\ref{eq:conductivity}) and (\ref{effectiveconcentration}).

\subsection{\label{step4} Scrutinizing the conductivity maximum positions}

\begin{figure}
\centering
\includegraphics[width=75mm]{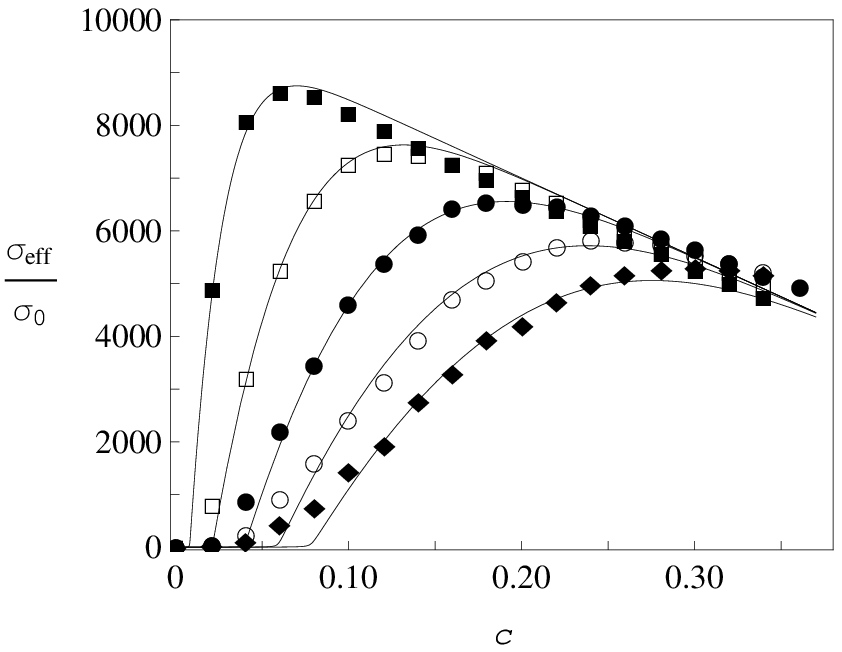}
\caption{\label{fig:Siekierski_HomogeneousLayers_t_fixed}
Simulation results \cite{Siekierski2005} for $\sigma_{\rm eff}$ as
a function of $c$ for a fixed shell thickness $t= 5\,{\rm \mu m}$,
constant shell conductivity, and  grain diameters $d = 3$
($\blacksquare$), 5 ($\square$), 7 ($\bullet$), 9 ($\circ$), and
$11\,{\rm \mu m}$ ($\blacklozenge$).  Solid lines: our theory
results for $\delta$'s given by Eq.~(\ref{eq:mapping}) with $K=
k$, $1.05\,k $, $1.05\,k $, $1.07\,k $, and $1.10\,k$,
respectively.}
\end{figure}

\begin{figure}
\centering
\includegraphics[width=75mm]{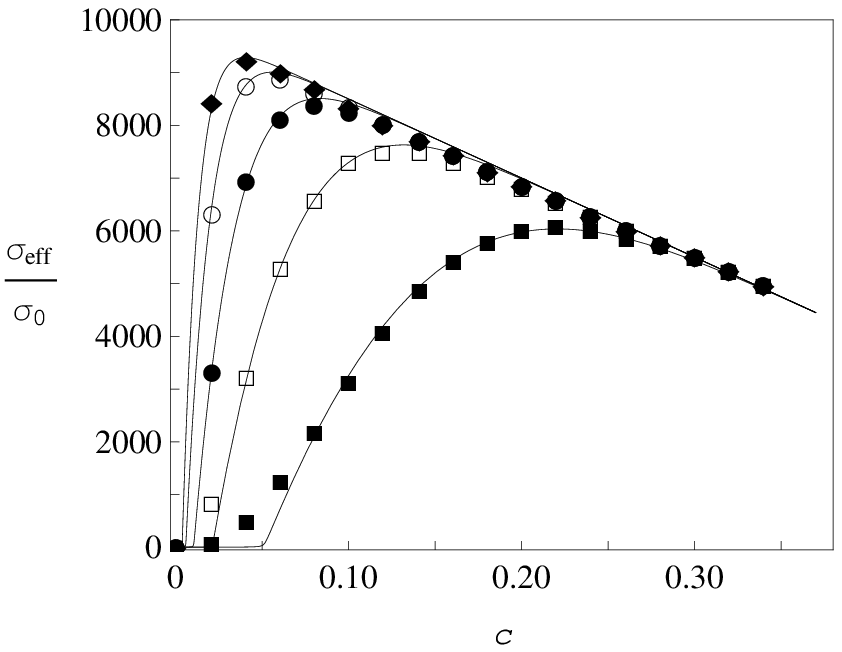}
\caption{\label{fig:Siekierski_HomogeneousLayers_d_fixed}
Simulation results \cite{Siekierski2005} for $\sigma_{\rm eff}$ as
a function of $c$ for a fixed grain diameter $d= 5\,{\rm \mu m}$,
constant shell conductivity, and shell thicknesses $t = 3$
($\blacksquare$), 5 ($\square$), 7 ($\bullet$), 9 ($\circ$), and
$11\,{\rm \mu m}$ ($\blacklozenge$). Solid lines: our theory
results for $\delta$'s given by Eq.~(\ref{eq:mapping}) with $K=
1.08\,k$, $1.05\,k $, $1.06\,k $, $1.07\,k $, and $1.06\,k$,
respectively.}
\end{figure}

\begin{figure}
\centering
\includegraphics[width=65mm]{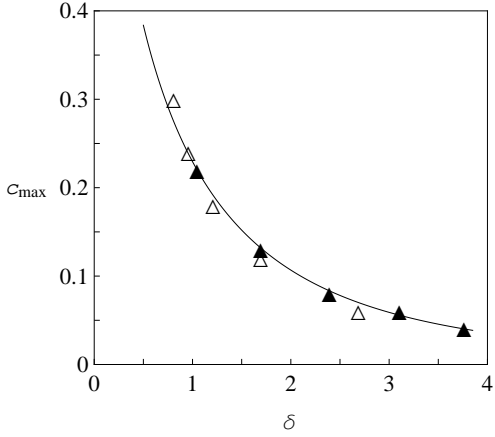}
\caption{\label{fig:Siekierski_MaximumLocation} $c_{\rm{max}}$ as
a function of $\delta$ according to Eqs.~(\ref{maximumLocation})
and (\ref{effectiveconcentration}) (solid line).  The empty
($\triangle$) and filled ($\blacktriangle$) triangles: the
$c_{\rm{max}}$ values recovered from simulation results
\cite{Siekierski2005} shown in
Figs.~\ref{fig:Siekierski_HomogeneousLayers_t_fixed} and
\ref{fig:Siekierski_HomogeneousLayers_d_fixed}, respectively; the
corresponding $\delta$ values were estimated by
Eq.~(\ref{eq:mapping}).}
\end{figure}

\begin{figure}
\centering
\includegraphics[width=75mm]{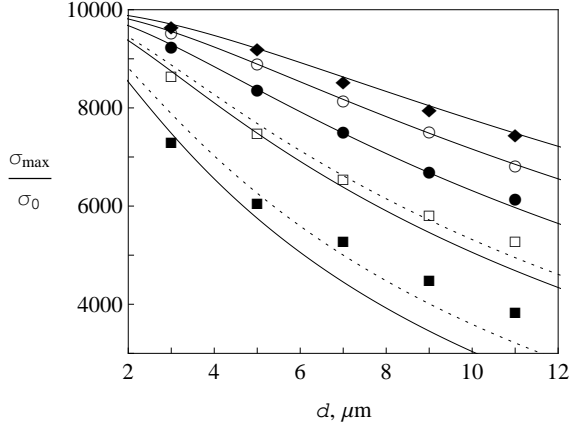}
\caption{\label{fig:Siekierski_MaxConductivity} Simulation results
\cite{Siekierski2005} for $\sigma_{\rm max}$ as a function of the
grain diameter $d$ for a constant shell conductivity and shell
thicknesses $ t = 3$ ($\blacksquare$), 5 ($\square$), 7
($\bullet$), 9 ($\circ$), and $11\,{\rm \mu m}$ ($\blacklozenge$).
Solid lines: our theory results for $\sigma_{\rm max}$ given by
Eqs.~(\ref{eq:CompositeElectrolyteconductivity}),
(\ref{maximumLocation}), and (\ref{eq:mapping}) with $K= k$.
Dotted lines: the same for $ t = 3$ and $5\,{\rm \mu m}$ at
$K=1.15\,k $ and $1.07\,k $, respectively.}
\end{figure}

Under the condition $\sigma_1\ll \sigma_0\ll\sigma_2$, typical of
simulations \cite{Siekierski2005,Siekierski2007},
Eq.~(\ref{eq:conductivity}) can be greatly simplified by passing
to the limit $\sigma_1 \to 0$ where it takes the form
\begin{eqnarray}
4\sigma_{\rm eff}^3  -2 \left[(2-3
\phi)\sigma_0-(1+3c-3\phi)\sigma_2\right] \sigma_{\rm eff}^2
\nonumber\\-(2-3c) \sigma_0\sigma_2\sigma_{\rm eff}=0.
\label{equationzeroSigma1}
\end{eqnarray}
A nontrivial physically meaningful solution to
Eq.~(\ref{equationzeroSigma1}) is
\begin{equation} \label{eq:CompositeElectrolyteconductivity}
\sigma_{\rm eff} = \frac{3}{4} \left( A+ \sqrt{B +A^2} \right),
\end{equation}
where \setcounter{equation}{48}
\begin{subequations}
\begin{equation} \label{eq:A}
 A\equiv  \left( \frac{2}{3} - \phi \right) \sigma_0 +
\left( \phi - c - \frac{1}{3} \right) \sigma_2,
\end{equation}
\begin{equation} \label{eq:B}
B\equiv \frac{4}{3} \left( \frac{2}{3} - c \right)
\sigma_0\sigma_2.
\end{equation}
\end{subequations}

For the data series in
Figs.~\ref{fig:Siekierski_HomogeneousLayers_t_fixed} and
\ref{fig:Siekierski_HomogeneousLayers_d_fixed},  the $\sigma_{\rm
eff}$ versus $c$ plots given by
Eqs.~(\ref{eq:CompositeElectrolyteconductivity}) and
Eq.~(\ref{eq:conductivity}) are indistinguishable.

The concentrations $c_{\rm max}$ where the conductivity maxima
occur are found from the conditions $\partial \sigma_{\rm
eff}/\partial c =0$ and $\partial^2 \sigma_{\rm eff}/\partial c^2
<0$. Since near these maxima $\sigma_{\rm eff} \gg \sigma_0$, it
follows from Eq.~(\ref{equationzeroSigma1}) and the first
condition that
\begin{equation}  \label{maximumLocation}
\left.{\partial \phi(c,\delta)}/{\partial c}\,\right|_{c = c_{\rm
{max}}} = 1,
\end{equation}
and that the derivatives ${\partial^2 \sigma_{\rm eff}}/{\partial
c^2}$ and ${\partial^2 \phi}/{\partial c^2}$ have the same sign at
$c = c_{\rm {max}}$. According to
Eq.~(\ref{effectiveconcentration}), ${\partial^2 \phi}/{\partial
c^2} <0$ for $\delta >0$. So, for such a $\delta$, the second
condition is fulfilled, and $\sigma_{\rm eff}$ has a local maximum
at $c_{\rm max}$ indeed. Its value $\sigma_{\rm max}$ is given by
Eqs.~(\ref{eq:CompositeElectrolyteconductivity}) at
$c=c_{\rm{max}}$ found from Eq.~(\ref{maximumLocation}).

The $c_{\rm {max}}$  versus   $\delta$ dependence given by
Eqs.~(\ref{maximumLocation})  and (\ref{effectiveconcentration})
is shown in Fig.~\ref{fig:Siekierski_MaximumLocation}. It agrees
very well with every pair of $c_{\rm {max}}$ and $\delta$ obtained
by processing simulation data \cite{Siekierski2005} with
Eqs.~(\ref{eq:conductivity}), (\ref{effectiveconcentration}), and
(\ref{eq:mapping}). This fact signifies the internal consistency
of our processing procedure. The  dependence of $\sigma_{\rm
{max}}$ on  the grain diameter $d$ (and, in fact, $\delta$) is
tested in Fig.~\ref{fig:Siekierski_MaxConductivity}. It is seen
that our theory reproduces almost the entire set of simulations
data \cite{Siekierski2005}. Noticeable discrepancies occur only
for smallest values of $\delta$ where the simulation errors are of
the greatest magnitude.

It is worthy of note that, provided  $\sigma_1\ll
\sigma_0\ll\sigma_2$, Eq.~(\ref{maximumLocation}) and the
inequality $\left.{\partial^2 \phi}/{\partial
c^2}\right|_{c=c_{\rm max}} <0 $ can be viewed as the second
derivative test for a local maximum of the shell volume
concentration $\phi(c,\delta) - c$. If the shells are penetrable,
this maximum was shown to occur at $c_{\rm{max}}$. In contrast,
there is no such a maximum, and therefore no local maximum for
$\sigma_{\rm {eff}}$ given by Eq.~(\ref{eq:conductivity}), in the
case of hard shells, where $\phi(c,\delta)$ is expressed by
Eq.~(\ref{hard}) (see
Fig.~\ref{fig:SiekierskiConductivity107a_4}). One way out of this
situation is based on the idea \cite{Nakamura1982,Nakamura1984} to
replace the conductivities $\sigma_q$ of the constituents by
certain conductivities $\sigma_q^*$ depending on not only
$\sigma_q$, but also an averaged property of the surrounding
medium in some form. In applications
\cite{Nan1993,Nan1991L,Nan1991} to CSEs, this approach is realized
in several steps: (1) introducing different quasi-two-phase models
of CSEs for the limiting cases of low and high values of $c$; (2)
calculating the effective conductivities in both limiting models
by standard one-particle methods; (3) sewing the solutions at some
characteristic concentration $v_2^*$, where the maximum of
$\sigma_{\rm {eff}}$ is observed. Evidently, $v_2^*$ serves as a
fitting parameter, and the dependence of $\sigma_{\rm {eff}}$ upon
$c$ reveals a nonphysical cusp at $c= v_2^*$ (see
\cite{Nan1993,Nan1991L,Nan1991}).

\subsection{\label{step5} Testing our model for the  case of inhomogeneous penetrable shells}

Now, simulation results \cite{Siekierski2006} for $\sigma_{\rm
eff}$ are processed with Eq.~(\ref{eq:nonuniformconductivity}).
The shell conductivity $\sigma^\prime_2$ was assumed in
\cite{Siekierski2006} to be distributed by a spherically-symmetric
Gaussian law, with a maximum of $\sigma^\prime_{\rm max}$ at the
distance $t/2$ from the surface of the grain and a minimum of
$\sigma^\prime_{\rm min}$ on the outer border of the shell (see
Table~\ref{tab:simulationparameters} for the numerical values).
The explicit expression for $\sigma^\prime_2 =\sigma^\prime_2(u)$
was not reported, and neither was the rule whereby a particular
conductivity value was assigned to each cubic cell belonging to
the shell.

Based on the above description, suppose that
\begin{equation}    \label{eq:GaussianSimulations}
\sigma^\prime_2(u)=\sigma^\prime_{\rm max} \exp
\left[-\frac{4\left(u-\delta^\prime/2\right)^2}{{\delta^\prime}^2}\ln\left(\frac{\sigma^\prime_{\rm
max}}{\sigma^\prime_{\rm min}} \right)\right].
\end{equation}
Let $n=t/a$ be the (even) number of the radially-distributed cubic
cells inside the shell, with their centers located at points
$u_p=(2p-1)\delta^\prime/2n$, $p=1, \dots, n$. If the conductivity
of the $p\,$th cell is defined as the value of $\sigma^\prime
_2(u)$ at $u_p$, then we can expect that the counterpart of the
distribution (\ref{eq:GaussianSimulations}) in our model has the
same functional form
\begin{equation}    \label{eq:Gaussian}
\sigma_2(u)=\sigma_{\rm max} \exp
\left[-\frac{4\left(u-\delta/2\right)^2}{\delta^2}\ln\left(\frac{\sigma_{\rm
max}}{\sigma_{\rm min}} \right)\right],
\end{equation}
with $\sigma_{\rm max}$ and $\sigma_{\rm min}$ equal to the
conductivities of the two central cells and the two outermost
cells, respectively:
$$\sigma_{\rm max} = \sigma^\prime_2(u_{n/2})=\sigma^\prime_2(u_{n/2+1})=
\sigma^\prime_{\rm max}\left(\frac{ \sigma^\prime_{\rm max}}{
\sigma^\prime_{\rm min}}\right)^{-1/n^2},$$
$$\sigma_{\rm min} = \sigma^\prime_2(u_{1})=\sigma^\prime_2(u_{n})=
\sigma^\prime_{\rm max}\left(\frac{ \sigma^\prime_{\rm max}}{
\sigma^\prime_{\rm min}}\right)^{-(n-1)^2/n^2}.$$ In the limit
$n\to \infty$, $\sigma_{\rm max} = \sigma_{\rm max}^\prime$ and
$\sigma_{\rm min} = \sigma_{\rm min}^\prime$; for any finite $n$,
$\sigma_{\rm max} < \sigma_{\rm max}^\prime$, $\sigma_{\rm min}
> \sigma_{\rm min}^\prime$, and
$$\frac{\sigma_{\rm max}}{\sigma_{\rm min}} = \left(\frac{ \sigma^\prime_{\rm max}}{
\sigma^\prime_{\rm min}}\right)^{(n-2)/n}.$$ These results
indicate that the conductivity parameters in the distribution
(\ref{eq:Gaussian}) are dependent of the details of simulations
\cite{Siekierski2006}. In the situation where these details are
unknown, one of these parameters, say, $\sigma_{\rm max}$, can be
treated as a fitting one.

Figures~\ref{fig:Siekierski_t_fixed} and
\ref{fig:Siekierski_d_fixed} demonstrate the results of processing
data \cite{Siekierski2006} by
Eq.~(\ref{eq:nonuniformconductivity}) with $\phi(c,\delta)$,
$\delta$, and $\sigma_2(u)$ given by
Eqs.~(\ref{effectiveconcentration}), (\ref{eq:mapping}), and
(\ref{eq:Gaussian}), respectively. The  used values of $K$ and
$\sigma_{\rm max}$ are summarized in
Tables~\ref{tab:numerical-inhomogeneous-1} and
\ref{tab:numerical-inhomogeneous-2}; for the sake of
simplification, it was taken $\sigma_{\rm min}=\sigma^\prime _{\rm
min}$. As seen, our model is capable of reproducing the simulation
data surprisingly well. Note also that according to the above
reasoning and for the given $\sigma^\prime_{\rm
max}/\sigma^\prime_{\rm min}=100$, ${\rm
log_{10}}\left(\sigma_{\rm max}/\sigma_{\rm min}\right) =2
(n-2)/n$. In the cases $t=9\, {\rm \mu m}$ ($n=18$) and $t=11\,
{\rm \mu m}$ ($n=22$), which are most appropriate for comparisons,
this relation gives ${\rm log_{10}}\left(\sigma_{\rm
max}/\sigma_{\rm min}\right) \approx 1.78$ and $1.82$,
respectively. The so-estimated values of $\sigma_{\rm
max}/\sigma_{\rm min}$ differ from those obtained by fitting by no
more than 17 and $12\%$.

\begin{figure}
\centering
\includegraphics[width=75mm]{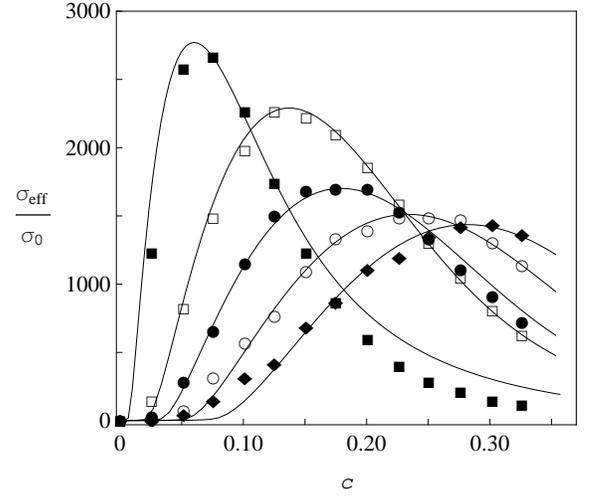}
\caption{\label{fig:Siekierski_t_fixed} Simulation results
\cite{Siekierski2006} for $\sigma_{\rm eff}$ as a function of $c$
for a fixed shell thickness $t= 5\,{\rm \mu m}$, grain diameters
$d = 3$ ($\blacksquare$), 5 ($\square$), 7 ($\bullet$), 9
($\circ$), and $11\,{\rm \mu m}$ ($\blacklozenge$), and Gaussian
shell conductivity profiles (\ref{eq:GaussianSimulations}).  Solid
lines: our theory results for the corresponding $\delta$'s and
shell conductivity profiles (\ref{eq:Gaussian}), found with the
mapping parameters listed in
Table~\ref{tab:numerical-inhomogeneous-1}. }
\end{figure}

\begin{figure}
\centering
\includegraphics[width=75mm]{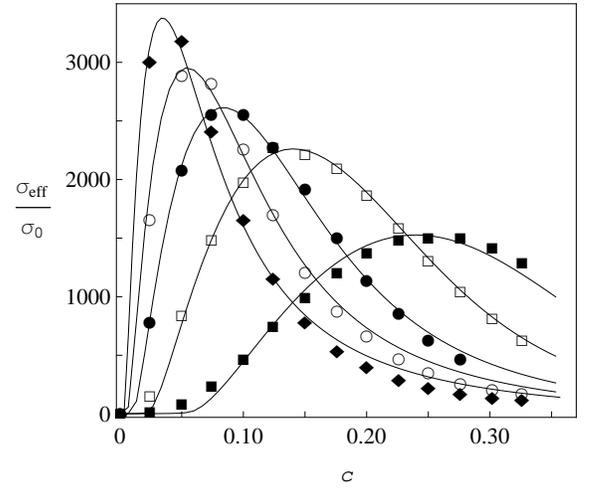}
\caption{\label{fig:Siekierski_d_fixed} Simulation results
\cite{Siekierski2006} for $\sigma_{\rm eff}$ as a function of $c$
for a fixed grain diameter $d= 5\,{\rm \mu m}$, shell thicknesses
$t = 3$ ($\blacksquare$), 5 ($\square$), 7 ($\bullet$), 9
($\circ$), and $11\,{\rm \mu m}$ ($\blacklozenge$), and Gaussian
shell conductivity profiles (\ref{eq:GaussianSimulations}). Solid
lines: our theory results for the corresponding $\delta$'s and
shell conductivity profiles (\ref{eq:Gaussian}), found with the
mapping parameters listed in
Table~\ref{tab:numerical-inhomogeneous-2}. }
\end{figure}

\begin{table}[tbh]
\caption{\label{tab:numerical-inhomogeneous-1} Parameters used to
fit the simulation data in Fig.~\ref{fig:Siekierski_t_fixed} by
Eq.~(\ref{eq:nonuniformconductivity}) with Gaussian shell
conductivity profiles (\ref{eq:Gaussian}) at $\sigma_{\rm
min}=\sigma_{\rm min}^\prime$; $\sigma_0 =10^{-8}\,{\rm{S/cm}}$,
$\sigma_1=10^{-12}\,{\rm{S/cm}}$.}
\begin{ruledtabular}
\begin{tabular}{lrrrrr}
$d,\, {\rm \mu m}$ & 3    & 5    & 7     & 9     & 11    \\
\hline
$K/k$    & 1.09 & 1.02 & 1.13  & 1.11  & 1.09   \\
${\rm log_{10}}\left(\sigma_{\rm max}/\sigma_{\rm min}\right)$
         & 1.83 & 1.89 & 1.82 & 1.88 & 1.98\\
\end{tabular}
\end{ruledtabular}
\end{table}

\begin{table}[tbh]
\caption{\label{tab:numerical-inhomogeneous-2} Parameters used to
fit the simulation data in Fig.~\ref{fig:Siekierski_d_fixed} by
Eq.~(\ref{eq:nonuniformconductivity}) with Gaussian shell
conductivity profiles (\ref{eq:Gaussian}) at $\sigma_{\rm
min}=\sigma_{\rm min}^\prime$; $\sigma_0 =10^{-8}\,{\rm{S/cm}}$,
$\sigma_1=10^{-12}\,{\rm{S/cm}}$.}
\begin{ruledtabular}
\begin{tabular}{lrrrrr}
$t,\, {\rm \mu m}$ & 3    & 5    & 7     & 9     & 11    \\
\hline
$K/k$    & 1.00  & 1.00  & 1.05 & 1.07 & 1.13  \\
${\rm log_{10}}\left(\sigma_{\rm max}/\sigma_{\rm min}\right)$
         & 1.90  & 1.89  & 1.85 & 1.85 & 1.87 \\
\end{tabular}
\end{ruledtabular}
\end{table}

\section{\label{sec:experiment} Application to experiment and discussion}

To exemplify the efficiency of the theory, we have applied it to
Liang's pioneering experimental data~\cite{Liang1973} for
$\sigma_{\rm eff}$ as a function of $c$ for real ${\rm
LiI/Al_2O_3}$ CSEs. The procedure involved several steps. First,
we processed data~\cite{Liang1973} with
Eq.~(\ref{eq:nonuniformconductivity}) assuming the hard cores to
be nonconductive (for alumina, $\sigma_1 \approx 1 \times
10^{-14}\,{\rm S/cm}$) and using the following three
approximations for the shell conductivity profile
$\sigma_2=\sigma_2(r)$:

(a) uniform shells ($0 < u < \delta_1$), $$x_2 = {\rm const};$$

(b) two-layer shells ($0 < u < \delta_2$),
\begin{equation} \label{eq:twoshellapproximation}
x_2(u)=\begin{cases} x_{2,1} & {\text{if} }
\quad \,\,0 < u < \delta_1,\\
x_{2,2} & {\text{if} }\quad  \delta_1<u <\delta_2;
\end{cases}
\end{equation}

(c) continuous shells of the sigmoid-type ($u > 0$),
\begin{equation}\label{eq:GeneratingFunction2}
x_2(u)= X_{2,1} + \frac{X_{2,2}-X_{2,1}}
{1+\exp\left(-\frac{u-\Delta_1}{\alpha}\right)}
+\frac{1-X_{2,2}}{1+\exp\left(-\frac{u-\Delta_2}{\alpha}\right)}.
\end{equation}
Here: $x_2= \sigma_2/\sigma_0$ and $x_{2,i} =
\sigma_{2,i}/\sigma_0$ are the relative conductivities, and
$\delta_1$ and $\delta_2$ are the relative thicknesses of the
layers; $X_{2,1}$, $X_{2,2}$, $\Delta_1$, and $\Delta_2$ are the
parameters of the generating function
(\ref{eq:GeneratingFunction2}). In the limit $\alpha\to 0$, where
the latter takes form (\ref{eq:twoshellapproximation}), they
become equal to the parameters $x_{2,1}$, $x_{2,2}$, $\delta_1$,
and $\delta_2$ of the two-layer model, respectively.

When processing the data with profile
(\ref{eq:GeneratingFunction2}), its parameters were varied so as
to smoothen it as much as possible; the parameter $\delta_M$ in
Eq.~(\ref{eq:nonuniformconductivity}) was fixed at a value of 5,
for a further increase of it did not affect the results.

The processing results are presented in
Figs.~\ref{fig:Liang_LiI-Al2O3-Processing},
\ref{fig:Liang_LiI-Al2O3-Profile} and Table~\ref{tab:Liang}. A
good agreement between the theory and experiment is achievable
under the condition that $\sigma_2(r)$ consists of two distinct
parts. To gain insight into this fact, we note that the use of
penetrable shells is a convenient way of modeling the effective
microstructure and conductivity of a system. Their $\sigma_2(r)$
is not equivalent to the actual conductivity distributions around
the hard cores, but is used to analyze these distributions and
possible mechanisms behind them.

Consider, for instance, model (\ref{eq:twoshellapproximation}),
which adequately describes the entire set of
data~\cite{Liang1973}. For this model,
Eq.~(\ref{eq:nonuniformconductivity}) can be represented as the
system of two equations
\begin{eqnarray}
\left[1-\phi(c, \delta_1)\right]\frac{\sigma_0^* -\sigma_{\rm
eff}}{2\sigma_{\rm eff}+\sigma_0^*} + c\,\frac{\sigma_1
-\sigma_{\rm
eff}}{2\sigma_{\rm eff}+\sigma_1} \nonumber \\
+\left[\phi(c, \delta_1)-c\right]\frac{\sigma_{2,1} -\sigma_{\rm
eff}}{2\sigma_{\rm eff}+\sigma_{2,1}}=0,
\label{eq:conductivityNewMatrix}
\end{eqnarray}
\begin{eqnarray}
&&(1 - \phi(c,\delta_1))\frac{\sigma_0^* - \sigma_{\rm
eff}}{2\sigma_{\rm eff} + \sigma_0^*}=(1 - \phi(c,\delta_2))
\frac{\sigma_0 - \sigma_{\rm eff}}{2\sigma_{\rm eff} + \sigma_0}
 \nonumber\\
&&+ (\phi(c,\delta_2) - \phi(c,\delta_1)) \frac{\sigma_{2,2} -
\sigma_{\rm eff}}{2\sigma_{\rm eff} + \sigma_{2,2}}.
\label{eq:sigma0-vs-c}
\end{eqnarray}
Under the conditions $c,\,\phi(c,\delta_1) \ll
\phi(c,\delta_2)<1$, the estimate $\sigma_{\rm eff} \approx
\sigma_0^*$ holds true [see Eq.~(\ref{eq:conductivityNewMatrix})],
$\sigma_{\rm eff}$ is basically  a function of $\sigma_0$, $c$,
$\delta_2$, and $\sigma_{2,2}$ [see Eq.~(\ref{eq:sigma0-vs-c})],
and so is $\sigma_0^*$.  In other words, at low $c$ and
$\phi(c,\delta_1)$, the electrical conduction in the system is
determined by $\sigma_0^*$, which is formed mainly by the
outermost part of $\sigma_2(r)$.

\begin{figure}
\centering
\includegraphics[width=75mm]{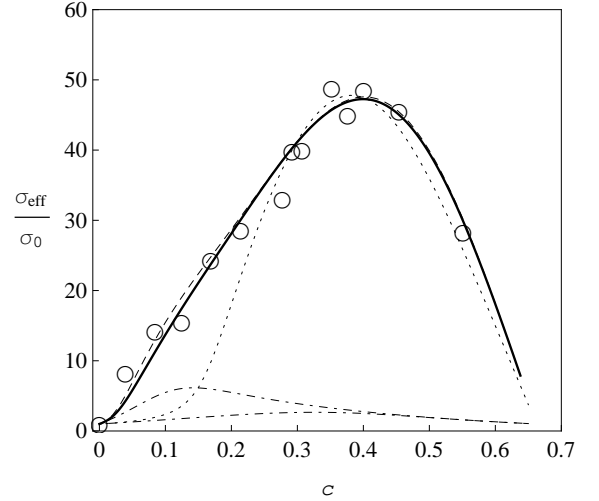}
\caption{\label{fig:Liang_LiI-Al2O3-Processing}  Experimental data
\cite{Liang1973} ($\circ$) for $\sigma_{\rm eff}$ of
$\rm{LiI/Al_2O_3}$ CSEs and their fits using (a) uniform shell
(\ref{eq:conductivity}) (dotted line), (b) two-layer
(\ref{eq:twoshellapproximation}) (dashed line), and (c)
sigmoid-type (\ref{eq:GeneratingFunction2}) (solid line)
approximations for $\sigma_2(r)$. The fitting parameters are
listed in Table \ref{tab:Liang}. The corresponding shell
conductivity profiles are shown in
Fig.~\ref{fig:Liang_LiI-Al2O3-Profile}.  The dotdashed lines
illustrate the predictions by the Maxwell-Garnett-type mixing
rules (\ref{eq:MGapproach}) for the same approximations (a) (lower
line) and (b) (upper line) and the same parameters, indicated  in
Table \ref{tab:Liang}.}
\end{figure}

\begin{figure}
\centering
\includegraphics[width=75mm]{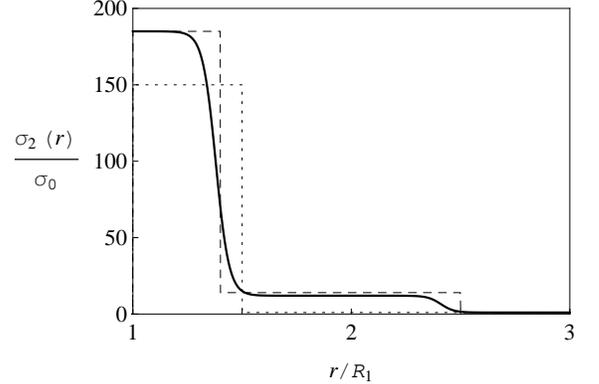}
\caption{\label{fig:Liang_LiI-Al2O3-Profile} The shell
conductivity profiles corresponding to the fits shown in
Fig.~\ref{fig:Liang_LiI-Al2O3-Processing}.}
\end{figure}

\begin{table}[tbh]
\caption{\label{tab:Liang} Parameters used to fit experimental
data \cite{Liang1973} for $\sigma_{\rm eff}$ of $\rm{LiI/Al_2O_3}$
CSEs using  (a) uniform shell (\ref{eq:conductivity}), (b)
two-layer (\ref{eq:twoshellapproximation}), and (c) sigmoid-type
(\ref{eq:GeneratingFunction2}) approximations for $\sigma_2(r)$;
$\sigma_0 = 2.5 \times 10^{-7}\,{\rm S/cm}$,  $x_1 = 0$.}
\begin{ruledtabular}
\begin{tabular}{llllll}
(a) & $x_2$ & $\delta$ & &  &  \\
     & 150    & 0.5   &  &  & \\
\hline
(b)  & $x_{2,1}$   &$x_{2,2}$ & $\delta_1$ & $\delta_2$ &  \\
     & 185         & 14  & 0.40       & 1.50       &  \\
\hline
(c)  & $X_{2,1}$    &$X_{2,2}$    & $\Delta_1$ & $\Delta_2$ &  $\alpha$ \\
     & 185         & 12        & 0.38       & 1.41       &   0.03    \\
\end{tabular}
\end{ruledtabular}
\end{table}

Physically, in view of Eq.~(\ref{eq:conductivityNewMatrix}),
$\sigma_0^*$ can be interpreted as the effective conductivity of
the host matrix in the CSE prepared by embedding filler particles
with hard cores, of radius $R_1$ and conductivity $\sigma_1$, and
penetrable shells, of relative thickness $\delta_1$ and
conductivity $\sigma_{\rm 2,1}$, into this matrix. The dependence
of $\sigma_0^*$ on $c$ can be recovered from
Eq.~(\ref{eq:sigma0-vs-c}) and, for ${\rm LiI/Al_2O_3}$ CSEs
\cite{Liang1973}, is shown in
Fig.~\ref{fig:Liang_LiI-Al2O3-Matrix}. For $c \lesssim 0.1$, it
very closely resembles the initial part of the $\sigma_{\rm eff}$
versus $c$ plot in Fig.~\ref{fig:Liang_LiI-Al2O3-Processing}. This
signifies that, despite being highly-conductive, the above shells
(inner layers in $\sigma_2(r)$) practically do not contribute to
$\sigma_{\rm eff}$ of ${\rm LiI/Al_2O_3}$ CSEs in this
concentration range.

Note that matrix processes enhancing the conductivity of the
matrix conductors in CSEs may include: a formation of defect-rich
space charge regions near the grain boundaries in a
polycrystalline matrix \cite{Maier1986}; development of a highly
conductive network of piled-up dislocations caused by
mechanically- and thermally-induced misfits
\cite{Dudney1987,Dudney1988,Muhlherr1988}; fast ionic transport
along matrix grain boundaries and/or dislocations
\cite{Phipps1981,Atkinson1988}; homogeneous doping of the matrix
through the dissolution of impurities and very fine particles in
it \cite{Wen1983,Dupree1983,Dudney1985}.

\begin{figure}
\centering
\includegraphics[width=75mm]{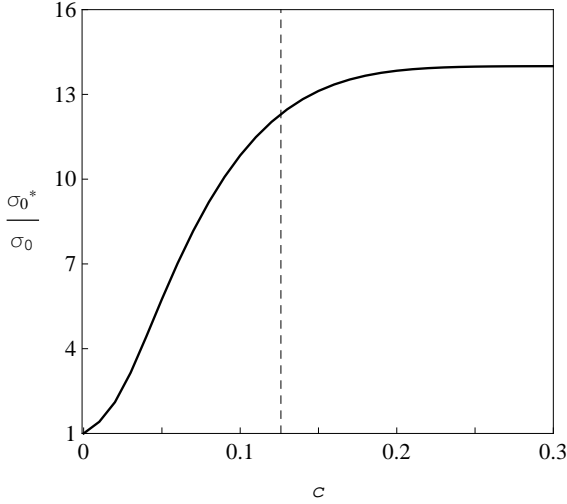}
\caption{\label{fig:Liang_LiI-Al2O3-Matrix} The matrix
conductivity $\sigma_0^*$ as a function of $c$ (solid line), as
recovered from Eq.~(\ref{eq:sigma0-vs-c}) for the two-shell
profile shown in Fig.~\ref{fig:Liang_LiI-Al2O3-Profile} (and its
parameters given in row (b) in Table~\ref{tab:Liang}). Dashed
line: the percolation threshold $c_{\rm c} \approx 0.126$ in the
system of the inner shells.}
\end{figure}

The situation changes drastically in the vicinity of the
percolation threshold  $c_{\rm c}$ for the indicated core-shell
particles. The value of $c_{\rm c}$ is found from the relation
$\phi(c_{\rm c},\delta_1) = {1}/{3}$ \cite{Sushko2013}, which for
$\delta_1 = 0.4$ gives $c_{\rm c} \approx 0.126$. It is the inner
part in $\sigma_2(r)$ that forms $\sigma_{\rm eff}$ at $c \gtrsim
c_{\rm c}$.

Typical examples of interfacial processes giving rise to highly
conductive regions around the filler particles are: a formation,
through preferential ion adsorption (desorption) at the
particle-matrix interface, of a space-charge layer enriched with
point defects \cite{Jow1979,Maier1984,Maier1985};  rapid ion
transport along the particle-matrix interface due to matrix
lattice distortions near it \cite{Phipps1981,Phipps1983};
stabilization of conductive non-equilibrium states  by the
adjacent filler particles \cite{Plocharski1988,Wieczorek1989};
formation of a new ``superstructure'' or interphase due to
chemical reactions at the interface \cite{Schmidt1988}. For ${\rm
LiI/Al_2O_3}$ CSEs, the inner part in $\sigma_2(r)$ can be
associated with a space charge layer. Indeed, our values
$\delta_1=0.4$ and $x_{2,1}=185$ correlate well with Jiang and
Wagner's estimates $\delta=0.4$ and $x_2= 324$
\cite{Jiang1995a,Jiang1995b} for the relative thickness and
relative conductivity of the space charge layer in ${\rm
LiI/Al_2O_3}$ CSEs modeled as cubic lattices with ideal random
distributions of cubic filler particles; estimates
\cite{Jiang1995a,Jiang1995b} were obtained by a method of
combination of a percolation model with the space charge layer
model.

For other types of composites, alternative physico-chemical
mechanisms are expected to come into play. In particular,
Eq.~(\ref{eq:nonuniformconductivity}) is
sufficient~\cite{Sushko2018polymer} to describe the observed
behavior of $\sigma_{\rm eff}$ for composite polymeric
electrolytes (CPEs) based on poly(ethy1ene oxide) (PEO) and
oxymethylene-linked PEO (OMPEO), provided the pertinent
$\sigma_2(r)$ consists of several parts. These account for: a
change of the matrix's conductivity in the course of preparation
of the composites (the outermost part in $\sigma_2(r)$);
amorphization of the polymer matrix by filler grains (the central
part); a stiffening effect of the filler on the amorphous phase
and effects caused by irregularities in the shape of the filler
grains (the innermost part).

It can be concluded from the above results that the functional
form (\ref{eq:nonuniformconductivity}) for $\sigma_{\rm eff}$ in
terms of the parameters of the hard-core-penetrable-shell model is
highly flexible and rather universal in the sense of being
applicable to various dispersed systems. At the same time, the
values of these parameters are not universal because  of a
diversity of physico-chemical mechanisms that not only form
$\sigma_{\rm eff}$ of real composite materials, but also alter the
properties of their constituents themselves. Consequently, these
values can be estimated provided that sufficiently extensive
experimental data are available.

The predictive power of the theory can be significantly increased
and, therefore, the amount of the required experimental work
considerable decreased by going beyond the limits of a pure
homogenization theory and employing certain model estimates for
the constituent's parameters and their dependences on various
factors, say, temperature. For instance, to recover the
temperature behavior of $\sigma_{\rm eff}$ for OMPEO-based CPEs
with different concentrations of polyacrylamide filler, it is
sufficient to use a 3-layer structure for $\sigma_2(r)$, assume
the conductivities of the layers and the matrix to obey, as
functions of temperature, the empirical three-parametric
Vogel-Tamman-Fulcher (VTF) equation, and recover the VTF
parameters for these conductivities by processing only three
conductivity isotherms for the CPEs. The reader is referred to
\cite{Sushko2018polymer} for the details.

\section{\label{sec:conclusion} Conclusion}
The main results of this paper are as follows:

(i) We give a self-consistent analytic solution to the problem of
the effective quasistatic electrical conductivity $\sigma_{\rm
eff}$ of a statistically homogeneous and isotropic dispersion of
hard-core--penetrable-shell particles with radially-symmetrical,
piecewise-continuous shell conductivity profile $\sigma_2(r)$; the
local conductivity value in the dispersion is assumed to be
determined by the distance from the point of interest to the
nearest particle. The solution effectively incorporates
many-particle effects in concentrated dispersions and is obtained
by: (a) generalizing the compact group approach
\cite{Sushko2007,Sushko2009CompGroups,Sushko2009AnisPart,Sushko2017}
to systems with complex-valued permittivities of the constituents;
(b) deriving the governing equation~(\ref{eq:equation}) for the
effective quasistatic complex permittivity $\hat{\varepsilon}_{\rm
eff}$ of the system; and (c) requiring that the boundary
condition \cite{Sillars1937} for the normal components of complex
electric fields in conducting dielectrics be satisfied. With the
latter requirement fulfilled, Eq.~(\ref{eq:equation}) becomes a
closed relation for $\hat{\varepsilon}_{\rm eff}$ in terms of the
statistical moments for the local deviations of the complex
permittivity distribution in the model dispersion from
$\hat{\varepsilon}_{\rm eff}$.

(ii) The desired $\sigma_{\rm eff}$, extracted from Eq. (26), is a
functional of the constituents' conductivities and volume
concentrations that obeys the integral
relation~(\ref{eq:nonuniformconductivity}). For the model under
consideration, this relation is expected to be rigorous in the
limit of static probing fields. The volume concentrations account
for the statistical microstructure of the system. They are
determined by statistical averages of products of the particles'
characteristic functions and can be estimated using other authors'
analytical \cite{Rikvold85,Torquato2013} and numerical
\cite{Lee1988,Rottereau2003} results for the volume concentration
of uniform shells.

(iii) The validity of the solution, at least for the parameter
values typical of CSEs and CPEs, is demonstrated by the results of
(a) mapping it onto extensive RRN simulation data
\cite{Siekierski2005,Siekierski2007,Siekierski2006} for the
composition and $\sigma_{\rm eff}$ of 3D dispersions comprising a
poorly-conductive, uniform matrix and isotropic particles with
nonconductive, hard cores and highly-conductive, fully-penetrable
shells with different $\sigma_2(r)$; and (b) applying it to
pioneering experimental data \cite{Liang1973} for real
LiI/Al$_2$O$_3$ CSEs.  The latter results also clarify the meaning
of $\sigma_2(r)$, reveal the role of its different parts in the
formation of $\sigma_{\rm eff}$, and indicate that both matrix and
interfacial processes contribute to enhanced electrical conduction
in LiI/Al$_2$O$_3$ CSEs.

To conclude, we note that Eqs.~(\ref{eq:equation}) and
(\ref{eq:nonuniformconductivity}) have already been shown to
efficiently describe electric percolation phenomena in random
composites \cite{Sushko2013}, electrical conductivity of
suspensions of insulating nanoparticles \cite{Sushko2016}, and
that of CPEs \cite{Sushko2018polymer}.

\section*{Acknowledgments}
We are deeply grateful to Prof. Luiz Roberto Evangelista and an
anonymous Referee for constructive remarks and recommendations on
improving the paper.


\end{document}